
\input harvmac
\input amssym
\noblackbox



\newfam\frakfam
\font\teneufm=eufm10
\font\seveneufm=eufm7
\font\fiveeufm=eufm5
\textfont\frakfam=\teneufm
\scriptfont\frakfam=\seveneufm
\scriptscriptfont\frakfam=\fiveeufm




\newfam\dsromfam
\font\tendsrom=dsrom10
\textfont\dsromfam=\tendsrom
\def\ds{\fam\dsromfam \tendsrom}


\newfam\mbffam
\font\tenmbf=cmmib10
\font\sevenmbf=cmmib7
\font\fivembf=cmmib5
\textfont\mbffam=\tenmbf
\scriptfont\mbffam=\sevenmbf
\scriptscriptfont\mbffam=\fivembf


\newfam\mbfcalfam
\font\tenmbfcal=cmbsy10
\font\sevenmbfcal=cmbsy7
\font\fivembfcal=cmbsy5
\textfont\mbfcalfam=\tenmbfcal
\scriptfont\mbfcalfam=\sevenmbfcal
\scriptscriptfont\mbfcalfam=\fivembfcal


\newfam\mscrfam
\font\tenmscr=rsfs10
\font\sevenmscr=rsfs7
\font\fivemscr=rsfs5
\textfont\mscrfam=\tenmscr
\scriptfont\mscrfam=\sevenmscr
\scriptscriptfont\mscrfam=\fivemscr




\def\tilde{\widetilde}

\def\hat{\widehat}

\def\bar{\overline}
\def\b{\bar}
\def\bsq#1{{{\b{#1}}^{\lower 2.5pt\hbox{$\scriptstyle 2$}}}}
\def\bexp#1#2{{{\b{#1}}^{\lower 2.5pt\hbox{$\scriptstyle #2$}}}}
\def\dotexp#1#2{{{#1}^{\lower 2.5pt\hbox{$\scriptstyle #2$}}}}


\def\rt2{\sqrt{2}}
\def\half {{1 \over 2}}

\def\det{\mathop{\rm det}}

\def\underrel#1\over#2{\mathrel{\mathop{\kern\z@#1}\limits_{#2}}}


\font\tenbifull=cmmib10
\font\tenbimed=cmmib7
\font\tenbismall=cmmib5
\textfont9=\tenbifull \scriptfont9=\tenbimed
\scriptscriptfont9=\tenbismall

\mathchardef\bbGamma="7000
\mathchardef\bbDelta="7001
\mathchardef\bbPhi="7002
\mathchardef\bbAlpha="7003
\mathchardef\bbXi="7004
\mathchardef\bbPi="7005
\mathchardef\bbSigma="7006
\mathchardef\bbUpsilon="7007
\mathchardef\bbTheta="7008
\mathchardef\bbPsi="7009
\mathchardef\bbOmega="700A
\mathchardef\bbalpha="710B
\mathchardef\bbbeta="710C
\mathchardef\bbgamma="710D
\mathchardef\bbdelta="710E
\mathchardef\bbepsilon="710F
\mathchardef\bbzeta="7110
\mathchardef\bbeta="7111
\mathchardef\bbtheta="7112
\mathchardef\bbiota="7113
\mathchardef\bbkappa="7114
\mathchardef\bblambda="7115
\mathchardef\bbmu="7116
\mathchardef\bbnu="7117
\mathchardef\bbxi="7118
\mathchardef\bbpi="7119
\mathchardef\bbrho="711A
\mathchardef\bbsigma="711B
\mathchardef\bbtau="711C
\mathchardef\bbupsilon="711D
\mathchardef\bbphi="711E
\mathchardef\bbchi="711F
\mathchardef\bbpsi="7120
\mathchardef\bbomega="7121
\mathchardef\bbvarepsilon="7122
\mathchardef\bbvartheta="7123
\mathchardef\bbvarpi="7124
\mathchardef\bbvarrho="7125
\mathchardef\bbvarsigma="7126
\mathchardef\bbvarphi="7127


\def\nablaslash{\not{\hbox{\kern-2pt $\nabla$}}}



\def\{{\lbrace}
\def\}{\rbrace}
\def\s{\sigma}

\def\CA{{\cal A}}
\def\CB{{\cal B}}

\def\CE{{\cal E}}

\def\CH{{\cal H}}

\def\CJ{{\cal J}}
\def\CK{{\cal K}}
\def\CL{{\cal L}}
\def\CM{{\cal M}}
\def\CN{{\cal N}}
\def\CO{{\cal O}}

\def\CR{{\cal R}}

\def\CT{{\cal T}}

\def\CW{{\cal W}}

\def\CY{{\cal Y}}


\def\1{{\ds 1}}

\def\a{\alpha}
\def\b{\beta}

\def\s{\sigma}

\def\p{\partial}


\def\{{\lbrace}
\def\}{\rbrace}

\def\a{\alpha}
\def\b{\beta}

\def\s{\sigma}

\def\p{\partial}

\font\sidenotefont=pxi at 6.5pt
\long\def\sidenote#1{%
  \vadjust{\llap{\smash{\vtop{%
    \parindent=0pt
    \hsize=0.7in
    \parfillskip=0pt
    \leftskip=0pt plus1fil
    \baselineskip=10pt\sidenotefont\vglue-\ht\strutbox #1}}\kern1em}}}


\lref\KapustinGMA{
  A.~Kapustin,
  ``Bosonic Topological Insulators and Paramagnets: a view from cobordisms,''
[arXiv:1404.6659 [cond-mat.str-el]].
}

\lref\WessCP{
  J.~Wess and J.~Bagger,
  ``Supersymmetry and supergravity,''
Princeton, USA: Univ. Pr. (1992) 259 p.
}

\lref\BanerjeeIZ{
  N.~Banerjee, J.~Bhattacharya, S.~Bhattacharyya, S.~Jain, S.~Minwalla and T.~Sharma,
  ``Constraints on Fluid Dynamics from Equilibrium Partition Functions,''
JHEP {\bf 1209}, 046 (2012).
[arXiv:1203.3544 [hep-th]].
}

\lref\LandsteinerSJA{
  K.~Landsteiner,
  ``Anomaly related transport of Weyl fermions for Weyl semi-metals,''
Phys.\ Rev.\ B {\bf 89}, 075124 (2014).
[arXiv:1306.4932 [hep-th]].
}

\lref\BachasNXA{
  C.~P.~Bachas, I.~Brunner, M.~R.~Douglas and L.~Rastelli,
 ``Calabi's diastasis as interface entropy,''
Phys.\ Rev.\ D {\bf 90}, no. 4, 045004 (2014).
[arXiv:1311.2202 [hep-th]].
}
\lref\LandsteinerKD{
  K.~Landsteiner, E.~Megias and F.~Pena-Benitez,
  ``Anomalous Transport from Kubo Formulae,''
Lect.\ Notes Phys.\  {\bf 871}, 433 (2013).
[arXiv:1207.5808 [hep-th]].
}

\lref\SeibergQD{
  N.~Seiberg,
  ``Modifying the Sum Over Topological Sectors and Constraints on Supergravity,''
JHEP {\bf 1007}, 070 (2010).
[arXiv:1005.0002 [hep-th]].
}

\lref\BardeenPM{
  W.~A.~Bardeen and B.~Zumino,
  ``Consistent and Covariant Anomalies in Gauge and Gravitational Theories,''
Nucl.\ Phys.\ B {\bf 244}, 421 (1984).
}

\lref\wip{
J.~Gomis, Z.~Komargodski, H.~Ooguri, N.~Seiberg and Y.~Wang,
work in progress.
}
\lref\JensenRGA{
  K.~Jensen, R.~Loganayagam and A.~Yarom,
  ``Chern-Simons terms from thermal circles and anomalies,''
[arXiv:1311.2935 [hep-th]].
}

\lref\JensenKKA{
  K.~Jensen, R.~Loganayagam and A.~Yarom,
  ``Anomaly inflow and thermal equilibrium,''
[arXiv:1310.7024 [hep-th]].
}

\lref\BanksZN{
  T.~Banks and N.~Seiberg,
  ``Symmetries and Strings in Field Theory and Gravity,''
Phys.\ Rev.\ D {\bf 83}, 084019 (2011).
[arXiv:1011.5120 [hep-th]].
}

\lref\ButterLTA{
  D.~Butter, B.~de Wit, S.~M.~Kuzenko and I.~Lodato,
  ``New higher-derivative invariants in N=2 supergravity and the Gauss-Bonnet term,''
JHEP {\bf 1312}, 062 (2013).
[arXiv:1307.6546 [hep-th].
}

\lref\Fradkinetal{
E.~S.~Fradkin and A.~A.~Tseytlin,
``Asymptotic Freedom In Extended Conformal Supergravities,''
Phys.\ Lett.\ B {\bf 110}, 117 (1982), ``One Loop Beta Function in Conformal Supergravities,''
Nucl.\ Phys.\ B {\bf 203}, 157 (1982); S.M. Paneitz, ``A quartic conformally
covariant differential operator for arbitrary peudo-Riemannian manifolds'',
arXiv:0803.4331; R.~J.~Riegert,
``A Nonlocal Action for the Trace Anomaly,''
Phys.\ Lett.\ B {\bf 134}, 56 (1984).}

\lref\JackEB{
  I.~Jack and H.~Osborn,
  ``Analogs for the $c$ Theorem for Four-dimensional Renormalizable Field Theories,''
Nucl.\ Phys.\ B {\bf 343}, 647 (1990).
}

\lref\OsbornGM{
  H.~Osborn,
  ``Weyl consistency conditions and a local renormalization group equation for general renormalizable field theories,''
Nucl.\ Phys.\ B {\bf 363}, 486 (1991).
}

\lref\JensenKJ{
  K.~Jensen, R.~Loganayagam and A.~Yarom,
  ``Thermodynamics, gravitational anomalies and cones,''
JHEP {\bf 1302}, 088 (2013).
[arXiv:1207.5824 [hep-th]].
}

\lref\GolkarKB{
  S.~Golkar and D.~T.~Son,
  ``Non-Renormalization of the Chiral Vortical Effect Coefficient,''
[arXiv:1207.5806 [hep-th]].
}

\lref\BershadskyCX{
  M.~Bershadsky, S.~Cecotti, H.~Ooguri and C.~Vafa,
  ``Kodaira-Spencer theory of gravity and exact results for quantum string amplitudes,''
Commun.\ Math.\ Phys.\  {\bf 165}, 311 (1994).
[hep-th/9309140].
}

\lref\DumitrescuHA{
  T.~T.~Dumitrescu, G.~Festuccia and N.~Seiberg,
  ``Exploring Curved Superspace,''
JHEP {\bf 1208}, 141 (2012).
[arXiv:1205.1115 [hep-th]].
}

\lref\SeibergVC{
  N.~Seiberg,
  ``Naturalness versus supersymmetric nonrenormalization theorems,''
Phys.\ Lett.\ B {\bf 318}, 469 (1993).
[hep-ph/9309335].
}

\lref\ClossetRU{
  C.~Closset, T.~T.~Dumitrescu, G.~Festuccia and Z.~Komargodski,
  ``Supersymmetric Field Theories on Three-Manifolds,''
JHEP {\bf 1305}, 017 (2013).
[arXiv:1212.3388 [hep-th]].
}

\lref\GrimmXP{
  R.~Grimm, M.~Sohnius and J.~Wess,
  ``Extended Supersymmetry and Gauge Theories,''
Nucl.\ Phys.\ B {\bf 133}, 275 (1978).
}

\lref\WessCP{
  J.~Wess and J.~Bagger,
  ``Supersymmetry and supergravity,''
Princeton, USA: Univ. Pr. (1992) 259 p.
}

\lref\KuzenkoUYA{
  S.~M.~Kuzenko, U.~Lindstrom, M.~Rocek, I.~Sachs and G.~Tartaglino-Mazzucchelli,
  ``Three-dimensional N=2 supergravity theories: From superspace to components,''
Phys.\ Rev.\ D {\bf 89}, 085028 (2014).
[arXiv:1312.4267 [hep-th]].
}

\lref\ClossetVG{
  C.~Closset, T.~T.~Dumitrescu, G.~Festuccia, Z.~Komargodski and N.~Seiberg,
  ``Contact Terms, Unitarity, and F-Maximization in Three-Dimensional Superconformal Theories,''
JHEP {\bf 1210}, 053 (2012).
[arXiv:1205.4142 [hep-th]].
}

\lref\HullJV{
  C.~M.~Hull and E.~Witten,
  ``Supersymmetric Sigma Models and the Heterotic String,''
Phys.\ Lett.\ B {\bf 160}, 398 (1985).
}

\lref\KomargodskiPC{
  Z.~Komargodski and N.~Seiberg,
  ``Comments on the Fayet-Iliopoulos Term in Field Theory and Supergravity,''
JHEP {\bf 0906}, 007 (2009).
[arXiv:0904.1159 [hep-th]].
}

\lref\WittenYC{
  E.~Witten,
  ``Phases of N=2 theories in two-dimensions,''
Nucl.\ Phys.\ B {\bf 403}, 159 (1993).
[hep-th/9301042].
}

\lref\JockersDK{
  H.~Jockers, V.~Kumar, J.~M.~Lapan, D.~R.~Morrison and M.~Romo,
  ``Two-Sphere Partition Functions and Gromov-Witten Invariants,''
Commun.\ Math.\ Phys.\  {\bf 325}, 1139 (2014).
[arXiv:1208.6244 [hep-th]].
}

\lref\DoroudXW{
  N.~Doroud, J.~Gomis, B.~Le Floch and S.~Lee,
  ``Exact Results in D=2 Supersymmetric Gauge Theories,''
JHEP {\bf 1305}, 093 (2013).
[arXiv:1206.2606 [hep-th]].
}

\lref\GomisWY{
  J.~Gomis and S.~Lee,
  ``Exact Kahler Potential from Gauge Theory and Mirror Symmetry,''
JHEP {\bf 1304}, 019 (2013).
[arXiv:1210.6022 [hep-th]].
}

\lref\BeniniUI{
  F.~Benini and S.~Cremonesi,
  ``Partition functions of $\CN=(2,2)$ gauge theories on $S^2$ and vortices,''
Commun.Math.Phys. (2014) July.
[arXiv:1206.2356 [hep-th]].
}

\lref\DoroudPKA{
  N.~Doroud and J.~Gomis,
  ``Gauge theory dynamics and K�hler potential for Calabi-Yau complex moduli,''
JHEP {\bf 1312}, 99 (2013).
[arXiv:1309.2305 [hep-th]].
}

\lref\DineBY{
  M.~Dine and N.~Seiberg,
  ``(2,0) Superspace,''
Phys.\ Lett.\ B {\bf 180}, 364 (1986).
}

\lref\ClossetVP{
  C.~Closset, T.~T.~Dumitrescu, G.~Festuccia, Z.~Komargodski and N.~Seiberg,
  ``Comments on Chern-Simons Contact Terms in Three Dimensions,''
JHEP {\bf 1209}, 091 (2012).
[arXiv:1206.5218 [hep-th]].
}

\lref\AharonyDHA{
  O.~Aharony, S.~S.~Razamat, N.~Seiberg and B.~Willett,
  ``3d dualities from 4d dualities,''
JHEP {\bf 1307}, 149 (2013).
[arXiv:1305.3924 [hep-th]].
}

\lref\DolanRP{
  F.~A.~H.~Dolan, V.~P.~Spiridonov and G.~S.~Vartanov,
  ``From 4d superconformal indices to 3d partition functions,''
Phys.\ Lett.\ B {\bf 704}, 234 (2011).
[arXiv:1104.1787 [hep-th]].
}

\lref\PeriwalMX{
  V.~Periwal and A.~Strominger,
  ``Kahler Geometry of the Space of $N=2$ Superconformal Field Theories,''
Phys.\ Lett.\ B {\bf 235}, 261 (1990).
}

\lref\SeibergPF{
  N.~Seiberg,
  ``Observations on the Moduli Space of Superconformal Field Theories,''
Nucl.\ Phys.\ B {\bf 303}, 286 (1988).
}

\lref\ImamuraUW{
  Y.~Imamura,
  ``Relation between the 4d superconformal index and the $S^3$ partition function,''
JHEP {\bf 1109}, 133 (2011).
[arXiv:1104.4482 [hep-th]].
}

\lref\FestucciaWS{
  G.~Festuccia and N.~Seiberg,
  ``Rigid Supersymmetric Theories in Curved Superspace,''
JHEP {\bf 1106}, 114 (2011).
[arXiv:1105.0689 [hep-th]].
}

\lref\CardyIE{
  J.~L.~Cardy,
  ``Operator Content of Two-Dimensional Conformally Invariant Theories,''
Nucl.\ Phys.\ B {\bf 270}, 186 (1986).
}

\lref\KomargodskiRB{
  Z.~Komargodski and N.~Seiberg,
  ``Comments on Supercurrent Multiplets, Supersymmetric Field Theories and Supergravity,''
JHEP {\bf 1007}, 017 (2010).
[arXiv:1002.2228 [hep-th]].
}

\lref\AsselTBA{
  B.~Assel, D.~Cassani and D.~Martelli,
  ``Supersymmetric counterterms from new minimal supergravity,''
JHEP {\bf 1411}, 135 (2014).
[arXiv:1410.6487 [hep-th]].
}

\lref\BaumeRLA{
  F.~Baume, B.~Keren-Zur, R.~Rattazzi and L.~Vitale,
  ``The local Callan-Symanzik equation: structure and applications,''
JHEP {\bf 1408}, 152 (2014).
[arXiv:1401.5983 [hep-th]].
}

\lref\WittenZZ{
  E.~Witten,
  ``Mirror manifolds and topological field theory,''
In *Yau, S.T. (ed.): Mirror symmetry I* 121-160.
[hep-th/9112056].
}

\lref\KinneyEJ{
  J.~Kinney, J.~M.~Maldacena, S.~Minwalla and S.~Raju,
  ``An Index for 4 dimensional super conformal theories,''
Commun.\ Math.\ Phys.\  {\bf 275}, 209 (2007).
[hep-th/0510251].
}

\lref\RomelsbergerEG{
  C.~Romelsberger,
  ``Counting chiral primaries in N = 1, d=4 superconformal field theories,''
Nucl.\ Phys.\ B {\bf 747}, 329 (2006).
[hep-th/0510060].
}

\lref\OsbornRNA{
  H.~Osborn and A.~Stergiou,
  ``Structures on the Conformal Manifold in Six Dimensional Theories,''
[arXiv:1501.01308 [hep-th]].
}

\lref\JackSHA{
  I.~Jack and H.~Osborn,
  ``Constraints on RG Flow for Four Dimensional Quantum Field Theories,''
Nucl.\ Phys.\ B {\bf 883}, 425 (2014).
[arXiv:1312.0428 [hep-th]].
}

\lref\WessYU{
  J.~Wess and B.~Zumino,
  ``Consequences of anomalous Ward identities,''
Phys.\ Lett.\ B {\bf 37}, 95 (1971).
}
\lref\GubserNZ{
  S.~S.~Gubser, I.~R.~Klebanov and A.~A.~Tseytlin,
  ``Coupling constant dependence in the thermodynamics of N=4 supersymmetric Yang-Mills theory,''
Nucl.\ Phys.\ B {\bf 534}, 202 (1998).
[hep-th/9805156].
}

\lref\SeibergPF{
  N.~Seiberg,
  ``Observations on the Moduli Space of Superconformal Field Theories,''
Nucl.\ Phys.\ B {\bf 303}, 286 (1988).
}

\lref\LandsteinerCP{
  K.~Landsteiner, E.~Megias and F.~Pena-Benitez,
  ``Gravitational Anomaly and Transport,''
Phys.\ Rev.\ Lett.\  {\bf 107}, 021601 (2011).
[arXiv:1103.5006 [hep-ph]].
}

\lref\DiFrancescoNK{
  P.~Di Francesco, P.~Mathieu and D.~Senechal,
  ``Conformal field theory,''
New York, USA: Springer (1997) 890 p.
}

\lref\LoganayagamPZ{
  R.~Loganayagam and P.~Surowka,
  ``Anomaly/Transport in an Ideal Weyl gas,''
JHEP {\bf 1204}, 097 (2012).
[arXiv:1201.2812 [hep-th]].
}

\lref\DumitrescuIU{
  T.~T.~Dumitrescu and N.~Seiberg,
  ``Supercurrents and Brane Currents in Diverse Dimensions,''
JHEP {\bf 1107}, 095 (2011).
[arXiv:1106.0031 [hep-th]].
}

\lref\BaggioVXA{
  M.~Baggio, V.~Niarchos and K.~Papadodimas,
  ``On exact correlation functions in SU(N) ${\cal N} = 2$ superconformal QCD,''
[arXiv:1508.03077 [hep-th]].
}

\lref\DumitrescuIU{
  T.~T.~Dumitrescu and N.~Seiberg,
  ``Supercurrents and Brane Currents in Diverse Dimensions,''
JHEP {\bf 1107}, 095 (2011).
[arXiv:1106.0031 [hep-th]].
}

\lref\GerchkovitzZRA{
  E.~Gerchkovitz,
  ``Constraints on the R-charges of free bound states from the R�melsberger index,''
JHEP {\bf 1407}, 071 (2014).
[arXiv:1311.0487 [hep-th]].
}

\lref\LandsteinerIQ{
  K.~Landsteiner, E.~Megias, L.~Melgar and F.~Pena-Benitez,
  ``Holographic Gravitational Anomaly and Chiral Vortical Effect,''
JHEP {\bf 1109}, 121 (2011).
[arXiv:1107.0368 [hep-th]].
}

\lref\ElitzurXJ{
  S.~Elitzur, Y.~Frishman, E.~Rabinovici and A.~Schwimmer,
  ``Origins of Global Anomalies in Quantum Mechanics,''
Nucl.\ Phys.\ B {\bf 273}, 93 (1986).
}

\lref\ClossetVRA{
  C.~Closset, T.~T.~Dumitrescu, G.~Festuccia and Z.~Komargodski,
  ``The Geometry of Supersymmetric Partition Functions,''
JHEP {\bf 1401}, 124 (2014).
[arXiv:1309.5876 [hep-th]].
}

\lref\WittenHU{
  E.~Witten and J.~Bagger,
  ``Quantization of Newton's Constant in Certain Supergravity Theories,''
Phys.\ Lett.\ B {\bf 115}, 202 (1982).
}

\lref\LouisDCA{
  J.~Louis, H.~Triendl and M.~Zagermann,
  ``N=4 Supersymmetric AdS5 Vacua and their Moduli Spaces,''
[arXiv:1507.01623 [hep-th]].
}

\lref\ClossetUDA{
  C.~Closset, T.~T.~Dumitrescu, G.~Festuccia and Z.~Komargodski,
  ``From Rigid Supersymmetry to Twisted Holomorphic Theories,''
Phys.\ Rev.\ D {\bf 90}, 085006 (2014).
[arXiv:1407.2598 [hep-th]].
}

\lref\KomargodskiRB{
  Z.~Komargodski and N.~Seiberg,
  ``Comments on Supercurrent Multiplets, Supersymmetric Field Theories and Supergravity,''
JHEP {\bf 1007}, 017 (2010).
[arXiv:1002.2228 [hep-th]].
}

\lref\DiPietroBCA{
  L.~Di Pietro and Z.~Komargodski,
  ``Cardy formulae for SUSY theories in $d =$ 4 and $d =$ 6,''
JHEP {\bf 1412}, 031 (2014).
[arXiv:1407.6061 [hep-th]].
}

\lref\DineBY{
  M.~Dine and N.~Seiberg,
  ``(2,0) Superspace,''
Phys.\ Lett.\ B {\bf 180}, 364 (1986).
}

\lref\NakayamaIS{
  Y.~Nakayama,
  ``Scale invariance vs conformal invariance,''
Phys.\ Rept.\  {\bf 569}, 1 (2015).
[arXiv:1302.0884 [hep-th]].
}

\lref\GrinsteinXBA{
  B.~Grinstein, D.~Stone, A.~Stergiou and M.~Zhong,
  ``Challenge to the $a$ Theorem in Six Dimensions,''
Phys.\ Rev.\ Lett.\  {\bf 113}, no. 23, 231602 (2014).
[arXiv:1406.3626 [hep-th]].
}

\lref\BrownSJ{
  L.~S.~Brown and J.~P.~Cassidy,
  ``Stress Tensors and their Trace Anomalies in Conformally Flat Space-Times,''
Phys.\ Rev.\ D {\bf 16}, 1712 (1977).
}

\lref\HoweBA{
  P.~S.~Howe and G.~Papadopoulos,
  ``N=2, D = 2 Supergeometry,''
Class.\ Quant.\ Grav.\  {\bf 4}, 11 (1987).
}

\lref\KomargodskiRB{
  Z.~Komargodski and N.~Seiberg,
  ``Comments on Supercurrent Multiplets, Supersymmetric Field Theories and Supergravity,''
JHEP {\bf 1007}, 017 (2010).
[arXiv:1002.2228 [hep-th]].
}

\lref\ClossetVG{
  C.~Closset, T.~T.~Dumitrescu, G.~Festuccia, Z.~Komargodski and N.~Seiberg,
  ``Contact Terms, Unitarity, and F-Maximization in Three-Dimensional Superconformal Theories,''
JHEP {\bf 1210}, 053 (2012).
[arXiv:1205.4142 [hep-th]].
}

\lref\KutasovXB{
  D.~Kutasov,
  ``Geometry on the Space of Conformal Field Theories and Contact Terms,''
Phys.\ Lett.\ B {\bf 220}, 153 (1989).
}

\lref\FriedanHI{
  D.~Friedan and A.~Konechny,
  ``Curvature formula for the space of 2-d conformal field theories,''
JHEP {\bf 1209}, 113 (2012).
[arXiv:1206.1749 [hep-th]].
}

\lref\NibbelinkWB{
  S.~Groot Nibbelink and L.~Horstmeyer,
  ``Super Weyl invariance: BPS equations from heterotic worldsheets,''
JHEP {\bf 1207}, 054 (2012).
[arXiv:1203.6827 [hep-th]].
}

\lref\deWitTN{
  B.~de Wit, J.~W.~van Holten and A.~Van Proeyen,
Nucl.\ Phys.\ B {\bf 184}, 77 (1981), erratum: Nucl.\ Phys.\ B {\bf 222}, 516 (1983).
}

\lref\BonoraCQ{
  L.~Bonora, P.~Pasti and M.~Bregola,
  ``Weyl Cocycles,''
Class.\ Quant.\ Grav.\  {\bf 3}, 635 (1986).
}

\lref\HerzogED{
  C.~P.~Herzog and K.~-W.~Huang,
  ``Stress Tensors from Trace Anomalies in Conformal Field Theories,''
Phys.\ Rev.\ D {\bf 87}, 081901 (2013).
[arXiv:1301.5002 [hep-th]].
}

\lref\Ginsparg{
  P.~H.~Ginsparg,
  ``Applied Conformal Field Theory,''
[hep-th/9108028].
}

\lref\OsbornRNA{
  H.~Osborn and A.~Stergiou,
  ``Structures on the Conformal Manifold in Six Dimensional Theories,''
JHEP {\bf 1504}, 157 (2015).
[arXiv:1501.01308 [hep-th]].
}

\lref\JackSHA{
  I.~Jack and H.~Osborn,
  ``Constraints on RG Flow for Four Dimensional Quantum Field Theories,''
Nucl.\ Phys.\ B {\bf 883}, 425 (2014).
[arXiv:1312.0428 [hep-th]].
}

\lref\GomisWOA{
  J.~Gomis and N.~Ishtiaque,
  ``Kahler Potential and Ambiguities in 4d N=2 SCFTs,''
[arXiv:1409.5325 [hep-th]].
}

\lref\ClossetVP{
  C.~Closset, T.~T.~Dumitrescu, G.~Festuccia, Z.~Komargodski and N.~Seiberg,
  ``Comments on Chern-Simons Contact Terms in Three Dimensions,''
JHEP {\bf 1209}, 091 (2012).
[arXiv:1206.5218 [hep-th]].
}

\lref\ColemanZI{
  S.~R.~Coleman and B.~R.~Hill,
  ``No More Corrections to the Topological Mass Term in QED in Three-Dimensions,''
Phys.\ Lett.\ B {\bf 159}, 184 (1985).
}

\lref\GerchkovitzGTA{
  E.~Gerchkovitz, J.~Gomis and Z.~Komargodski,
  ``Sphere Partition Functions and the Zamolodchikov Metric,''
JHEP {\bf 1411}, 001 (2014).
[arXiv:1405.7271 [hep-th]].
}

\lref\GrinsteinCKA{
  B.~Grinstein, A.~Stergiou and D.~Stone,
  ``Consequences of Weyl Consistency Conditions,''
JHEP {\bf 1311}, 195 (2013).
[arXiv:1308.1096 [hep-th]].
}

\lref\GrisaruDM{
  M.~T.~Grisaru and M.~E.~Wehlau,
  ``Prepotentials for (2,2) supergravity,''
Int.\ J.\ Mod.\ Phys.\ A {\bf 10}, 753 (1995).
[hep-th/9409043].
}

\lref\GrisaruDR{
  M.~T.~Grisaru and M.~E.~Wehlau,
  ``Superspace measures, invariant actions, and component projection formulae for (2,2) supergravity,''
Nucl.\ Phys.\ B {\bf 457}, 219 (1995).
[hep-th/9508139].
}

\lref\GatesDU{
  S.~J.~Gates, Jr., M.~T.~Grisaru and M.~E.~Wehlau,
  ``A Study of general 2-D, N=2 matter coupled to supergravity in superspace,''
Nucl.\ Phys.\ B {\bf 460}, 579 (1996).
[hep-th/9509021].
}

\lref\NakayamaWDA{
  Y.~Nakayama,
   ``Consistency of local renormalization group in d=3,''
Nucl.\ Phys.\ B {\bf 879}, 37 (2014).
[arXiv:1307.8048 [hep-th]].
}

\lref\deWitZA{
  B.~de Wit, S.~Katmadas and M.~van Zalk,
  ``New supersymmetric higher-derivative couplings: Full N=2 superspace does not count!,''
JHEP {\bf 1101}, 007 (2011).
[arXiv:1010.2150 [hep-th]].
}

\lref\Polchinski{
  J.~Polchinski,
  ``String theory. Vol. 1: An introduction to the bosonic string,''
}

\lref\KuzenkoGVA{
  S.~M.~Kuzenko,
  ``Super-Weyl anomalies in N=2 supergravity and (non)local effective actions,''
JHEP {\bf 1310}, 151 (2013).
[arXiv:1307.7586].
}

\lref\Buican{
  M.~Buican and T.~Nishinaka,
  ``Compact Conformal Manifolds,''
JHEP {\bf 1501}, 112 (2015).
[arXiv:1410.3006 [hep-th]].
}

\lref\LorenzenPNA{
  J.~Lorenzen and D.~Martelli,
  ``Comments on the Casimir energy in supersymmetric field theories,''
[arXiv:1412.7463 [hep-th]].
}

\lref\ClossetPDA{
  C.~Closset and S.~Cremonesi,
  ``Comments on $ \CN  = (2, 2)$ supersymmetry on two-manifolds,''
JHEP {\bf 1407}, 075 (2014).
[arXiv:1404.2636 [hep-th]].
}

\lref\WittenXJ{
  E.~Witten,
  ``Topological Sigma Models,''
Commun.\ Math.\ Phys.\  {\bf 118}, 411 (1988).
}

\lref\KutasovSV{
  D.~Kutasov and N.~Seiberg,
  ``Number of degrees of freedom, density of states and tachyons in string theory and CFT,''
Nucl.\ Phys.\ B {\bf 358}, 600 (1991).
}

\lref\ZamolodchikovGT{
  A.~B.~Zamolodchikov,
  ``Irreversibility of the Flux of the Renormalization Group in a 2D Field Theory,''
JETP Lett.\  {\bf 43}, 730 (1986), [Pisma Zh.\ Eksp.\ Teor.\ Fiz.\  {\bf 43}, 565 (1986)].
}

\lref\BeniniNOA{
  F.~Benini and A.~Zaffaroni,
 ``A topologically twisted index for three-dimensional supersymmetric theories,''
JHEP {\bf 1507}, 127 (2015).
[arXiv:1504.03698 [hep-th]].
}

\lref\DeserYX{
  S.~Deser and A.~Schwimmer,
  ``Geometric classification of conformal anomalies in arbitrary dimensions,''
Phys.\ Lett.\ B {\bf 309}, 279 (1993).
[hep-th/9302047].
}

\lref\KetovES{
  S.~V.~Ketov,
  ``2-d, N=2 and N=4 supergravity and the Liouville theory in superspace,''
Phys.\ Lett.\ B {\bf 377}, 48 (1996).
[hep-th/9602038].
}

\lref\BaggioIOA{
  M.~Baggio, V.~Niarchos and K.~Papadodimas,
  ``tt$^{*}$ equations, localization and exact chiral rings in 4d $\CN $ =2 SCFTs,''
JHEP {\bf 1502}, 122 (2015).
[arXiv:1409.4212 [hep-th]].
}
\lref\KetovTB{
  S.~V.~Ketov and S.~O.~Moch,
Class.\ Quant.\ Grav.\  {\bf 11}, 11 (1994).
[hep-th/9306140].
}
\lref\OsbornGM{
  H.~Osborn,
  ``Weyl consistency conditions and a local renormalization group equation for general renormalizable field theories,''
Nucl.\ Phys.\ B {\bf 363}, 486 (1991).
}

\lref\NekrasovWG{
  N.~A.~Nekrasov,
  ``Lectures on curved beta-gamma systems, pure spinors, and anomalies,''
[hep-th/0511008].
}

\lref\Plesser{
  M.~Bertolini, I.~V.~Melnikov and M.~R.~Plesser,
  ``Accidents in (0,2) Landau-Ginzburg theories,''
JHEP {\bf 1412}, 157 (2014).
[arXiv:1405.4266 [hep-th]].
}

\lref\AuzziYIA{
  R.~Auzzi and B.~Keren-Zur,
 ``Superspace formulation of the local RG equation,''
JHEP {\bf 1505}, 150 (2015).
[arXiv:1502.05962 [hep-th]].
}

\lref\TachikawaTQ{
  Y.~Tachikawa,
  ``Five-dimensional supergravity dual of a-maximization,''
Nucl.\ Phys.\ B {\bf 733}, 188 (2006).
[hep-th/0507057].
}

\lref\GreenDA{
  D.~Green, Z.~Komargodski, N.~Seiberg, Y.~Tachikawa and B.~Wecht,
  ``Exactly Marginal Deformations and Global Symmetries,''
JHEP {\bf 1006}, 106 (2010).
[arXiv:1005.3546 [hep-th]].
}

\lref\BaggioSNA{
  M.~Baggio, V.~Niarchos and K.~Papadodimas,
  ``Exact correlation functions in $SU(2)$ $\CN=2$ superconformal QCD,''
Phys.\ Rev.\ Lett.\  {\bf 113}, no. 25, 251601 (2014).
[arXiv:1409.4217 [hep-th]].
}

\lref\Naka{
  Y.~Nakayama,
  ``Local renormalization group functions from quantum renormalization group and holographic bulk locality,''
JHEP {\bf 1506}, 092 (2015).
[arXiv:1502.07049 [hep-th]].
}

\lref\ClossetRNA{
  C.~Closset, S.~Cremonesi and D.~S.~Park,
  ``The equivariant A-twist and gauged linear sigma models on the two-sphere,''
JHEP {\bf 1506}, 076 (2015).
[arXiv:1504.06308 [hep-th]].
}


\rightline{WIS/06/14-JUL-DPPA}
\vskip-50pt
\Title{
} {\vbox{\centerline{Anomalies, Conformal Manifolds, and Spheres}
 }}

\vskip-15pt
\centerline{Jaume Gomis,${}^1$ Po-Shen Hsin,${}^2$ Zohar Komargodski,${}^3$  Adam Schwimmer,${}^3$}
\centerline{Nathan Seiberg,${}^4$ and Stefan Theisen${}^5$}
\vskip15pt
  \centerline{\it ${}^1$ Perimeter Institute for Theoretical Physics, Waterloo, Ontario, N2L 2Y5, Canada}
\centerline{\it ${}^2$ Department of Physics, Princeton University, Princeton, NJ 08544}
  \centerline{\it ${}^3$ Weizmann Institute of Science, Rehovot 76100, Israel}
\centerline{\it ${}^4$ School of Natural Sciences, Institute for Advanced Study, Princeton, NJ 08540, USA}
\centerline{\it ${}^5$ Max-Planck-Institut f\"ur Gravitationsphysik, Albert-Einstein-Institut, 14476 Golm, Germany}
\vskip25pt

\noindent
The two-point function of exactly marginal operators leads to a universal contribution to the trace anomaly in even dimensions. We study  aspects of this trace anomaly, emphasizing its interpretation as a sigma model, whose target space $\CM$ is the space of conformal field theories (a.k.a.\ the conformal manifold). When the underlying quantum field theory is supersymmetric, this sigma model has to be appropriately supersymmetrized. As examples, we consider in some detail $\CN=(2,2)$ and $\CN=(0,2)$ supersymmetric theories in $d=2$ and $\CN=2$ supersymmetric theories in $d=4$. This reasoning leads to new information about the conformal manifolds of these theories, for example, we show that the manifold is K\"ahler-Hodge and we further argue that it has vanishing K\"ahler class. For $\CN=(2,2)$ theories in $d=2$ and $\CN=2$ theories in $d=4$ we also show that the relation between the sphere partition function and the K\"ahler potential of $\CM$ follows immediately from the appropriate sigma models that we construct.  Along the way we find several examples of potential trace anomalies that obey the Wess-Zumino consistency conditions, but can be ruled out by a more detailed analysis.

\bigskip
\Date{September 2015}

\newsec{Introduction}

Some $d$-dimensional conformal field theories have exactly marginal operators $\{\CO_I\}$.  This means that when we add them to the action with coupling constants $\lambda^I$,
\eqn\deformc{\delta S= {1\over\pi^{d/2}}\sum_I \int d^d x \lambda^I \CO_I(x)~}
the theory remains conformal.  The coefficients $\lambda^I$ parameterize the space of conformal field theories, a.k.a.\ the conformal manifold $\CM$.
The two-point functions
\eqn\twopt{\langle \CO_I(x)\CO_J(y)\rangle={g_{IJ}(\lambda^K)\over (x-y)^{2d}}~}
define a metric, known as the Zamolodchikov metric~\ZamolodchikovGT.  It is the metric on the conformal manifold \SeibergPF.  It carries
nontrivial information that cannot be removed by redefinitions of the coupling constants $\lambda^I$~\KutasovXB.
For example, the Ricci scalar associated to $g_{IJ}(\lambda^K)$ is invariant under all
such redefinitions.

The purpose of this note is to explore the geometry and the topology of $\CM$.\foot{In worldsheet string theory, $\CM$ can be interpreted as the space of classical vacua of the theory. In the AdS$_{d+1}$/CFT$_d$ correspondence, the conformal manifold of the CFT$_d$ is interpreted as the space of vacua in AdS$_{d+1}$ (see e.g.~\TachikawaTQ). These correspondences allow to connect our results to various other topics. }
Our main tool will be the conformal anomaly first discussed in~\OsbornGM.
By allowing $\lambda^I$ to be spacetime dependent background fields, \OsbornGM\ derived
a contribution to the trace of the energy-momentum tensor, which depends on the
Zamolodchikov metric $g_{IJ}$.

In supersymmetric theories the anomaly above must be supersymmetrized.  This introduces a few new elements into the analysis. First, it leads to restrictions on the local form of the metric  $g_{IJ}$ and it also leads to global restrictions. Second, the anomaly forces us to introduce some contact terms. We will study both aspects in detail.

In section 2 we review the analysis of these conformal anomalies (without supersymmetry).  Here we spell out the conditions they have to satisfy and show how a careful analysis leads to new constraints beyond the Wess-Zumino consistency conditions~\WessYU.

In the remaining sections we will study $\CN=(2,2)$ and $\CN=(0,2)$ theories in two dimensions and $\CN=2$ theories in four dimensions.

Our discussion of two-dimensional $\CN=(2,2)$ theories in section 3 leads to a new proof\foot{In this paper we assume that the coupling constants can be promoted  to $\CN=(2,2)$ chiral and twisted chiral superfields.  This assumption is non-trivial as it can fail in some cases \wip.}  that  $\CM$ factorizes into a space $\CM_c$ depending on chiral couplings ($\lambda$, $\bar \lambda$) and a space $\CM_{tc}$   depending on twisted chiral couplings ($\tilde \lambda$, $\bar{\tilde \lambda}$). Their K\"ahler potentials are $K_c(\lambda,\bar\lambda)$ and $K_{tc}(\tilde \lambda,\bar{\tilde\lambda})$.  We will show that $\CM$ must be Hodge\foot{A K\"ahler-Hodge manifold is a K\"ahler manifold for which the flux of the K\"ahler two-form through any two-cycle is an integer.} and will further argue that its K\"ahler class should be trivial. This, in particular, shows that $\CM$ cannot be a smooth compact manifold.

We will also study the sphere partition function.  Without supersymmetry, there are   counterterms that render it ambiguous.  With $\CN=(2,2)$ supersymmetry there are two ways, denoted $A$ and $V$, to place the theory on the sphere and the partition function has universal content.   It is given by
\eqn\spherett{Z_A=\left({r\over r_0}\right)^{c\over3} e^{-K_c(\lambda,\bar\lambda)}
\qquad ; \qquad Z_V=\left({r\over r_0}\right)^{c\over3} e^{-K_{tc}(\tilde \lambda,\bar{\tilde\lambda})} ~.}
Here $r$ is the radius of the sphere and  $r_0$ a (scheme dependent) scale.
The dependence on $r$ reflects the ordinary conformal anomaly.
As we will show, the appearance of $K_c$ or $K_{tc}$ reflects another contribution
to the conformal anomaly depending on exactly marginal couplings.
The identifications \spherett\ were conjectured in \refs{\JockersDK} and proven
in \refs{\GomisWY} as well as in~\GerchkovitzGTA\  based on the work
of \refs{\BeniniUI\DoroudXW-\DoroudPKA}.

In section 4 we will discuss ${\cal N}=(0,2)$ theories in $d=2$.  Our analysis
leads to restrictions on the metric on $\CM$ and shows that $\CM$ is Hodge and suggests that its
K\"ahler class is trivial.  But we will argue that the
two-sphere partition function is not universal.

Section 5 is devoted to $\CN=2$ theories in $d=4$.  Again, the sphere partition
function has universal content and computes the K\"ahler potential on $\CM$
\eqn\spherefourd{
Z=\left({r\over r_0}\right)^{-4a}\,e^{K/12}\,.}
This relation was proven in~\GerchkovitzGTA\ as well as in~\GomisWOA\ and was further
used in \refs{\BaggioIOA\BaggioSNA-\BaggioVXA}.
$\CN=2$ supersymmetry fixes an additional contribution to
the conformal anomaly depending on a four-tensor in $\CM$
in terms of the Riemann tensor of the Zamolodchikov metric.
As in two dimensions, our analysis shows that $\CM$ is Hodge and suggests that its
K\"ahler class is trivial.\foot{We do not make this claim for $\CN=1$ theories in four dimensions. In fact,~\Buican\ suggested a construction of compact conformal manifolds in $\CN=1$ theories.}

Our discussion is reminiscent of that of \refs{\ClossetVG,\ClossetVP}.  In both cases nontrivial contact
terms are identified.  They cannot be absorbed by supersymmetric local counterterms and therefore
correspond to anomalies.  They reflect short distance physics and can be analyzed in the
flat space theory.  Then, these contact terms have interesting consequences when the
theory is placed on the sphere.

For a related supersymmetric analysis of  conformal anomalies in $\CN=1$ theories in
$d=4$ and $\CN=2$ theories  in $ d=3$ see \refs{\AuzziYIA,\NakayamaWDA}.

Four appendices contain technical results. Appendix A concerns with the normalization of
the anomalies. In Appendix B we collect some properties of the
Fradkin-Tseytlin-Paneitz-Riegert  (FTPR) operator, which appears in the anomaly in
$d=4$.
Appendix C reviews $(2,2)$ and $(0,2)$ supersymmetry in two dimensions and their linearized supergravities. Appendix D considers $(2,2)$ Poincar\'e supergravity in superconformal gauge  (which always exists locally). We classify the allowed  rigid supersymmetric backgrounds in this gauge.

\newsec{The anomaly associated with the metric on $\CM$}

In momentum space the two-point functions \twopt\ take the following form
\eqn\momspace{\langle \CO_I(p)\CO_J(-p)\rangle\sim g_{IJ}
\biggl\{ \matrix{ p^{d}
& d=2n+1\cr
p^{2n}\log\left({\Lambda^2\over p^2}\right) & d=2n~.}}
The explicit scale (or cutoff) $\Lambda$ in the logarithm does not violate scale
invariance. The reason is that rescaling
$\Lambda$ changes the answer by a polynomial in $p^2$, which is a contact term.  The correlation function at separated points~\twopt\
therefore remains intact under rescaling $\Lambda$. Such logarithms appear abundantly in conformal field
theories (CFTs). Even though they do not violate the conformal Ward identities,
they lead to anomalies (i.e.\ the non-conservation of the dilatation charge in the
presence of non-vanishing background fields). One way to detect it is to make the couplings, $\lambda^I$, $x$-dependent.
Then, the trace anomaly in even dimensions includes a term, roughly of the form \refs{\OsbornGM,\FriedanHI}:\foot{This term is in addition to the ordinary conformal anomalies, which depend only on the spacetime metric.}
\eqn\anomalypoint{T_\mu^\mu\supset  g_{IJ}\lambda^I\square^{d\over 2}\lambda^J~.}
The precise action of the Laplacian could be to the left and to the right. We will specify
this later for $d=2$ and $d=4$.

We study   CFTs with spacetime metric $\gamma_{\mu\nu}$ and spacetime dependent coupling constants $\lambda^I$.  We assume that the theory can be regulated in a diffeomorphism-invariant fashion.  Specifically, we assume that the energy-momentum tensor is conserved even at coincident points (apart from the ordinary Ward identity relations). We will be interested in the partition function $Z[\gamma_{\mu\nu};\lambda^I]$ and its variation
$\delta_\sigma \log Z$ under infinitesimal Weyl transformations
\eqn\gammato{\delta_\sigma \gamma_{\mu\nu} =2\,\delta\sigma\, \gamma_{\mu \nu}~}
with infinitesimal $\delta \sigma$ of compact support.
Naively, conformal invariance means that the variation vanishes.  But because of the anomaly, it does not.  This variation satisfies a number of important properties:
\item{1.} $Z[\gamma_{\mu\nu};\lambda^I]$
is a nonlocal functional of its arguments.  However, its variation $\delta_\sigma \log Z$
is a {\it local} functional of $\gamma_{\mu\nu}$, $\lambda^I$ and $\delta \sigma$.
\item{2.} $\delta_\sigma \log Z$ must be coordinate invariant in spacetime.
\item{3.} It must be coordinate invariant in $\CM$.  Below we will argue that it should also be globally well defined on $\CM$.
\item{4.} It must obey the Wess-Zumino consistency condition~\WessYU
\eqn\WZcond{
\delta_{\sigma_1}\delta_{\sigma_2}\log Z-\delta_{\sigma_2}\delta_{\sigma_1}\log Z=0\,.}
\item{5.} A term in $\delta_\sigma \log Z$ that is obtained by a Weyl variation of a local term is considered trivial.  An anomaly is a ``cohomologically nontrivial'' term.  It cannot be removed by changing a counterterm.  Equivalently, it cannot  be removed by changing  the renormalization scheme~\BonoraCQ. Therefore, even though the anomaly arises due to a short distance regulator, it is universal -- it does not depend on the regularization.

\medskip
Let us start in $d=2$. The infinitesimal Weyl variation of $\log Z$ responsible for the
trace anomaly \anomalypoint\ and the ordinary trace anomaly
is given by
\eqn\OsEq{\delta_\sigma \log Z={c\over24\pi} \int d^2 x\, \delta \sigma\sqrt \gamma R-{1\over4\pi}\int d^2x\,\delta \sigma\sqrt \gamma \, g_{IJ}
\gamma^{\mu\nu}\del_\mu \lambda^I\del_\nu\lambda^J~.}
Here $R$ is the Ricci scalar and the first term is the universal contribution due to the   central charge~$c$.  The normalization of the second term is worked out in Appendix A.

The anomaly functional~\OsEq\ includes a sigma model with target space $\CM$.
It manifestly obeys the Wess-Zumino consistency condition because it is Weyl invariant.   There is no  local counterterm, whose Weyl variation
yields~\OsEq.  Therefore, \OsEq\ is cohomologically nontrivial.
In the language of~\DeserYX\ one could
refer to the first term in~\OsEq\ as a type-A and to the second term as a
type-B anomaly.\foot{A type-A
anomaly vanishes for $x$-independent $\delta\sigma$ when the background fields
have trivial topology. A type-B anomaly
does not vanish for constant sigma even for trivial topology and reflects
logarithms in certain correlation functions.}

An important part of our discussion will be the analysis of the allowed local counterterms (related to item 5 in the list above). In two dimensions, an important counterterm is
\eqn\countertwo{\int d^2x\,\sqrt\gamma \,R F(\lambda^I) ~.}
Its Weyl variation is
\eqn\countertwov{\delta_\sigma \int d^2x\,\sqrt\gamma \,R F(\lambda^I) =-2\int d^2x\,\sqrt \gamma\, \square \delta\sigma F(\lambda^I) ~.}
We will see various consequences of this counterterm in what follows.

In addition to~\OsEq, there are other potential trace anomalies that we need to consider.
First, we have the parity odd type-B anomaly
\eqn\OsEqNP{\delta_\sigma \log Z\supset\int d^2x\,\delta \sigma\sqrt \gamma \ B_{IJ}\epsilon^{\mu\nu}\del_\mu \lambda^I\del_\nu\lambda^J~,}
with an anti-symmetric two-form $B_{IJ}$ on the conformal manifold $\CM$.
Also, we have two  type-A anomalies
\eqn\AnomVect{\delta_\sigma \log Z\supset\int d^2x\,\partial_\mu \delta \sigma\sqrt \gamma \, V_{I}\gamma^{\mu\nu}\del_\nu \lambda^I~,}
and
  \eqn\AnomVecti{\delta_\sigma \log Z\supset\int d^2x\,\partial_\mu \delta \sigma\sqrt \gamma \, \tilde V_{I}\epsilon^{\mu\nu}\del_\nu \lambda^I~.}
They are characterized by one-forms $V_I$ and $\tilde V_I$  on $\CM$.  We note that \AnomVecti\ is invariant under the gauge transformation $\tilde V_I \to \tilde V_I + \partial_I \tilde f$. On the other hand, \AnomVect\ transforms under $ V_I \to  V_I + \partial_I f$, but the change is cohomologically trivial.  It can be absorbed in the Weyl variation of the local counterterm \countertwo\ with $F\sim f(\lambda^I) $. This allows us to identify $V_I$ and $\tilde V_I$ as connections on the conformal manifold $\CM$.

We will now show that
even though~\OsEqNP,\AnomVect,\AnomVecti\ obey the Wess-Zumino consistency conditions,
a more detailed analysis leads to  further restrictions, ruling out these anomalies.
This demonstrates that  constraints that go beyond the standard
cohomological analysis can further restrict  anomalies.

First, a simple argument excludes all  type-B anomalies   that are beyond that in~\OsEq. We   recall that a type-B anomaly is associated to a logarithm appearing in a correlation function. Without loss of generality we can study the theory in flat Euclidean spacetime with $\gamma_{\mu\nu}=\delta_{\mu \nu}$.
 Consider the momentum space correlation function of the exactly marginal operators
\eqn\correc{\langle O_{I_1}(p_1)O_{I_2}(p_2) \cdots O_{I_n}(p_n)\rangle =  \log \Lambda\ \delta\left(\sum p_r\right) A_{I_1I_2\cdots I_n}+\cdots~,}
where the ellipses on the right-hand side represent terms independent of the UV cutoff $\Lambda$.  Since the operators $O_I$ are exactly marginal, the coefficient of the logarithm must be ultra-local, i.e.\ a polynomial in momentum (otherwise, there would be a beta function for the couplings $\lambda^I$). Scale invariance constrains $A_{I_1I_2\cdots I_n}$ to be quadratic polynomials in the momenta $p_r$.  Therefore, we can determine $A_{I_1I_2\cdots I_n}$ by picking specific simple combinations of momenta.  For example, for $p_2=-p_1$ and $p_3=p_4=\cdots =0$ it is clear that $A_{I_1I_2\cdots I_n}\sim p_1^2 \partial_{I_3}\partial_{I_4}\cdots \partial_{I_n}g_{I_1I_2}$. Similar other specific cases show that $A_{I_1I_2\cdots I_n}$ is determined entirely by derivatives of the Zamolodchikov metric. This means that the additional parity odd anomaly~\OsEqNP\ controlled by a two-form cannot be present.  More precisely, the argument above shows that $H_{IJK}=\del_{[I}B_{JK]}=0$ and hence $B_{IJ}$ is locally given by $B_{IJ}=\del_{[I}B_{J]}$ for some one-form $B_{J}$. By integration by parts we find that this anomaly is now identical to the type-A anomaly~\AnomVecti. We will discuss it below.

Next, we argue that $V_I$ and $\tilde V_I$ in the type-A anomalies \AnomVect,\AnomVecti\
must satisfy $\del_{[I}V_{J]}=\del_{[I}\tilde V_{J]}=0$, i.e. these connections are
flat. These anomalies can be extracted from the following energy-momentum correlator
\eqn\TOO{\langle T_{\mu\nu}(z)\CO_I(x)\CO_J(y)\rangle  ~.}
Using the conformal Ward identity at separated points,  the correlator must be proportional to $g_{IJ}$. There could also be contributions with support at $x=y\neq z$, or $z=x\neq y$, or $z=y\neq x$. In the first case the only  contact term allowed by dimensional analysis contains $T_\mu^\mu$, which has zero separated-points correlation functions. In the second and third case, we can have the contact term $T_{\mu\nu}(x)\CO_I(0)\sim \delta_{\mu\nu}\delta^{(2)}(x)\CM_{I}^{K}\CO_K(0)$ with some matrix $\CM_I^K$. This would lead to a logarithmic term in the three-point function~\TOO, of the type already analyzed above, and hence, it is proportional to the Zamolodchikov metric and does not contribute to the anomalies $V_I$ and $\tilde V_I$.

Therefore, all the terms in~\TOO\ that are associated to separated points physics are proportional to $g_{IJ}$.  They cannot lead to nonzero ``field strengths'' $\del_{[I}V_{J]}$ or $\del_{[I}\tilde V_{J]}$, which are anti-symmetric in $I$ and $J$. Thus, at least locally, the connections $V_I$ and $\tilde V_I$ are pure gauge and the associated anomalies vanish.

Summarizing, we have seen that even though~\OsEq,\AnomVect,\AnomVecti, are a priori allowed anomalies that obey the Wess-Zumino consistency conditions, they can be all excluded. This will have important consequences in what follows.

Thus far we limited ourselves to deformations by exactly marginal operators with coefficients $\lambda^I$  as in~\deformc.  If the CFT also has conserved currents $j_\mu^a$, then it is natural to couple them to classical background fields $A^a_\mu$ and examine the anomaly as a function of these fields. The anomaly sigma models now depend on the spacetime metric $\gamma_{\mu\nu}$, the couplings $\lambda^I$, and the gauge fields $A_\mu^a$.  The operators $\CO_I$ are taken to carry charges $-q^a_I$ and the coupling constants $\lambda^I$ thus carry charges $q^a_I$. Away from $\lambda^I=0$ some of the symmetries generated by $j_\mu^a$ may be thus explicitly broken.
Related expressions appear in \FriedanHI.  In addition to the previous requirements of conformal invariance, coordinate invariance in spacetime and on $\CM$, we should now also demand  gauge invariance. The equation~\OsEq\ is modified by simply replacing $\del_\mu\lambda^I\rightarrow\nabla_\mu \lambda^I=\del_\mu\lambda^I-iq^a_JA_\mu^a\lambda^J $. In addition, one could encounter new anomalies that contain the field strength $F_{\mu\nu}^a$. There could also be 't Hooft anomalies under gauge transformations.

In $d=4$ the local functional  that reproduces the
logarithm in the two-point function~\momspace~ is a four-derivative local term. One can construct it by starting with the
ansatz $\delta_\sigma \log Z\supset \int d^4x\, \delta \sigma \sqrt \gamma \, g_{IJ} \square \lambda^I\square \lambda^J+\cdots $ and covariantize this expression both in spacetime and in $\CM$. One also requires that  it satisfies the Wess-Zumino consistency condition~\WZcond. After some work\foot{We use the
convention $[\nabla_\mu,\nabla_\nu]V_\rho=R_{\mu\nu\rho}{}^\sigma V_\sigma$.}
one finds the expression\foot{The fact that the anomaly
$\int d^4x\,\sqrt\gamma\,\delta \sigma\,
g_{IJ}\del_\nu \lambda^J\del_\mu \lambda^IR^{\mu\nu}$ is proportional to the Zamolodchikov metric at the fixed point plays a very important role in perturbative proofs of the strong version of the $a$-theorem~\refs{\JackEB,\OsbornGM,\GrinsteinCKA\JackSHA-\BaumeRLA}. The situation in $d=6$ unfortunately appears to be more complicated~\refs{\GrinsteinXBA,\OsbornRNA}. For a review see~\NakayamaIS.}
\eqn\OsEqfourd{\delta_\sigma\log Z\supset
{1\over192\pi^2}\int d^4x\,\sqrt\gamma\,\delta \sigma\left(g_{IJ} \hat\square \lambda^I\hat
\square \lambda^J-2\,g_{IJ} \del_\mu \lambda^I\left(R^{\mu\nu}-{1\over 3}\gamma^{\mu\nu} R \right)
\del_\nu \lambda^J\right)~.}
Above $\hat\square\lambda^I=\square\lambda^I
+\Gamma^{I}_{JK}\del^\mu \lambda^J\del_\mu\lambda^K$, where $\Gamma^{I}_{JK}$ is the
usual Christoffel connection on $\CM$. The ordinary Laplacian $\square$ is enriched to $\hat \square$ so that the anomaly is coordinate invariant on ${\cal M}$, as we demand in general.
At this juncture $\Gamma^I_{JK}$ could be an arbitrary connection, not necessarily the Levi-Civita one. However, demanding  that \OsEqfourd\ satisfies the Wess-Zumino consistency condition forces $\Gamma^I_{JK}$ to be the Levi-Civita connection. Note that~\OsEqfourd\ coincides with expressions that appeared in~\JackSHA\
and~\OsbornRNA\ in related contexts. The combination~\OsEqfourd\ can be viewed as an interesting variant of the
Fradkin-Tseytlin-Paneitz-Riegert operator  \Fradkinetal, which we discuss further in
Appendix B.

While we do not present an exhaustive classification of anomalies in four dimensions,
there is an additional conformal anomaly that depends on a
four-tensor on $\CM$ with components $c_{IJKL}$ that we would like to mention:
\eqn\addano{\delta_\sigma \log Z
\supset\int d^4x\,\delta \s \sqrt{\gamma}\ c_{IJKL}\del_\mu\lambda^I\del^\mu\lambda^J
\del_\nu\lambda^K\del^\nu \lambda^L~.}
The four-tensor $c_{IJKL}$ may be either an independent rank-four tensor on the manifold
$\CM$, or it may be fixed by the Zamolodchikov metric, e.g.\ $c_{IJKL}\sim g_{IJ}g_{KL}$
or $c_{IJKL}\sim R_{IKJL}+R_{JKIL}$, where $R_{IKJL}$ is the Riemann
tensor on ${\cal M}$.\foot{We thank Y.~Nakayama for a discussion on the topic and for stressing the potential relevance of the anomaly~\addano\ to the question of locality in AdS$_{d+1}$. See for instance~\Naka.} The Wess-Zumino consistency condition~\WZcond\ does not imply a relation between $c_{IJKL}$ and the Zamolodchikov metric.  However, in section 5 we will show that in $\CN=2$ supersymmetric theories, such a relation must exist, and  $c_{IJKL}$ is proportional to the Riemann curvature tensor of the Zamolodchikov metric.

For future references, let us also list some of the allowed counterterms in four dimensions
\eqn\counterfour{\int d^4x\, \sqrt \gamma\,\Big(R^2\ F_1(\lambda^I)+R_{\mu\nu}^2\ F_2(\lambda^I)+R_{\mu\nu\rho\sigma}^2\ F_3(\lambda^I)  + \cdots \Big)  ~.}

We will be particularly interested in the case where the underlying theory is supersymmetric.  Then, the exactly marginal couplings $\lambda^I$ reside in various superfields \SeibergVC.  If the superconformal field theory (SCFT) can be regularized in a supersymmetric manner, then we must further require that the local anomaly functionals above be  supersymmetrized.  We will study some of the consequences of supersymmetrizing~\OsEq\ and \OsEqfourd.  We will show that the remaining ambiguity \countertwo\ in the renormalization scheme in two dimensions and \counterfour\ in four dimensions is restricted to have holomorphic dependence on the coupling constants.  This fact has several important consequences.  In particular, it makes the sphere partition function meaningful (up to a K\"ahler transformation generated by a holomorphic function~\refs{\GerchkovitzGTA,\GomisWOA}).

\newsec{$\CN=(2,2)$ Supersymmetry in $d=2$}

Our goal in this section is to determine  the conformal anomaly and analyze its consequences in $\CN=(2,2)$ supersymmetric theories in two dimensions.
Here the exactly marginal parameters belong either to background chiral multiplets or twisted chiral multiplets, which we denote by $\lambda^I$ and $\tilde\lambda^A$, respectively.

First, we should supersymmetrize the anomaly \OsEq\ and the counterterm \countertwo.  For that we need to place the theory not only in curved space but in curved superspace \FestucciaWS.  $\CN=(2,2)$ supergravity was discussed in~\refs{\HoweBA\GrisaruDM\GrisaruDR\GatesDU-\ClossetPDA} and, in particular, the possibilities for rigid supersymmetry in curved space were analyzed in \ClossetPDA. (We repeat this analysis in the superconformal gauge in Appendix D.)

We should discuss two distinct supergravity formulations known as $U(1)_V$ and $U(1)_A$ supergravities~\HoweBA.  These are labeled by whether the $U(1)$ symmetry preserved in the Poincar\'e supergravity theory is vector or axial.\foot{To follow the discussion below (in our    analysis of two-dimensional theories) no familiarity with supergravity is necessary.}

In terms of the $(2,2)$ SCFT this distinction is the following.  The $(2,2)$ SCFT has a $U(1)_V\times U(1)_A$ R-symmetry.  We can couple either $U(1)_V$ or $U(1)_A$ to a background gauge field, but an anomaly prevents us from coupling both of them to background fields.  Correspondingly, the coincident points divergences and the associated contact terms can preserve either $U(1)_V$ or $U(1)_A$ R-symmetry but not both.  These contact terms are described by the corresponding supergravity.  Equivalently, we assume that the theory can be regularized while preserving   diffeomorphism invariance and supersymmetry as well as either $U(1)_V$ or $U(1)_A$. In particular, we assume that there are no gravitational anomalies so that $c_L=c_R$.

We find it convenient to use a simplification specific to two dimensions.  Since locally every two-dimensional metric is conformally flat, we can describe the metric using the conformal factor $\sigma$ -- the Liouville field.  This statement is easily supersymmetrized.  Every supergravity background can be described locally by a superconformal factor in a superfield. In $U(1)_A$ supergravity it is in a chiral superfield $\Sigma$ and in $U(1)_V$ supergravity it is in a twisted chiral superfield $\tilde \Sigma$ (see Appendix $C$).  The corresponding superconformal variations, whose anomalies we are interested in, are $\delta \Sigma$ and $\delta \tilde \Sigma$ respectively.  In what follows we will concentrate mainly on $U(1)_A$.  It is straightforward to repeat it for $U(1)_V$.

The supersymmetrization of the conformal anomaly \OsEq\ is then straightforward.
In the regularization preserving $U(1)_A$, the anomaly is given by
\eqn\actiontwotwoA{
\delta_{\Sigma}\log Z_A = -{c\over 24\pi} \int d^2 x d^4\theta(\delta \Sigma + \delta \bar \Sigma)(\Sigma + \bar \Sigma)+
{1\over 4\pi}\int d^2 x\,d^4\theta\,\Big(\delta \Sigma \, {\cal K}(\lambda,\bar\lambda,\tilde\lambda,\bar{\tilde\lambda})+ c.c.\Big)
~,}
where $\CK$ is a complex function of the exactly marginal couplings.\foot{In the full supergravity without using the conformal gauge the anomaly takes the form
\eqn\actiontwotwob{
\delta_{\Sigma}\log Z_A
=-{c\over24\pi}\left(\int d^2x\,d^2\theta\, {\cal E}\, \CR\,\delta\Sigma+\hbox{c.c.}\right)+ {1\over 4\pi}\int d^2 x\,d^4\theta\,E\Big(\delta \Sigma\,{\cal K}\, (\lambda,\bar\lambda,\tilde\lambda,\bar{\tilde\lambda}) + c.c.\Big)}
where $\CE$ is the chiral superspace measure,   $E$ is the Berezinian superfield, and here $\CR$ is a chiral superfield that contains the Ricci scalar in its $\theta^2$ component. The first term represents the ordinary anomaly.}
Clearly, these expressions obey the Wess-Zumino consistency condition.

One might try to integrate \actiontwotwoA\ to find the $\Sigma$ dependence of $\log Z_A$.  Although this can be done as a local expression in terms of $\Sigma$, the answer is nonlocal.  The point is that it is valid and local in the superconformal gauge, but it is nonlocal in other gauges.  This property makes it particularly interesting, as it cannot be absorbed in local counterterms.

In order to proceed we must find the most general supersymmetric expression, which is local in any gauge, and can serve as a local counterterm.  This is the supersymmetrization of \countertwo.  In $U(1)_A$ the local counterterm is~\GerchkovitzGTA\
\eqn\counterterm{S_A={1\over 4\pi}\int d^2x\,d^2\theta\,\CR\, F(\lambda)+\hbox{c.c.}= {1\over 4\pi}\int d^2x\,d^4\theta\, \bar \Sigma\, F(\lambda)+\hbox{c.c.}~,}
where $\CR= \bar D^2\bar \Sigma$ is the chiral curvature superfield in superconformal gauge.  The counterterm \counterterm\ depends only on the chiral parameters $\lambda$ and the dependence is holomorphic.\foot{In the full supergravity without using the conformal gauge the counterterm is
\eqn\countertermf{
S_A={1\over 4\pi}\int d^2x\,d^2\theta\, {\cal E}\, \CR\, F(\lambda)+\hbox{c.c.}~.}
}
Under a super-Weyl transformation
\eqn\superweylA{
\delta_{\Sigma}S_A= {1\over 4\pi}\int d^2x\,d^4\theta\, \left(\,\delta \bar\Sigma\, F(\lambda)+\delta \Sigma\, \bar F(\bar\lambda)\right) ~.}

Further restrictions on $\CK$ can be found
by expanding~\actiontwotwoA\ in components and requiring that the forbidden two-dimensional anomalies~\OsEqNP,\AnomVect,\AnomVecti\ are absent.  Note that this goes beyond the Wess-Zumino consistency conditions. After some algebra, the conclusion is that $\CK$ is real and
\eqn\CKsum{\CK= K_c(\lambda,\bar \lambda) - K_{tc}(\tilde \lambda, \bar {\tilde \lambda}) ~,}
and therefore the metric on $\CM$ is a product metric of two K\"ahler manifolds~$\CM=\CM_c\times \CM_{tc}$. The K\"ahler potential on $\CM_c$ is $K_c$ and it depends only on the chiral parameters and the K\"ahler potential on $\CM_{tc}$ is $K_{tc}$ and it depends only on the twisted chiral parameters.  This splitting between the chiral and the twisted chiral parameters is well known and is natural in the context of type II string theory, where
$(2,2)$ worldsheet theories lead to $\CN=2$ supersymmetry in spacetime. The hypermultiplet and the vector multiplet metrics are factorized as a consequence.  Here we see that it follows from properties of anomalies on the worldsheet.

We conclude that the anomaly is
\eqn\actiontwotwoAc{
\delta_{\Sigma}\log Z_A= -{c\over24\pi} \int d^2 x d^4\theta (\delta \Sigma + \delta \bar \Sigma)(\Sigma + \bar \Sigma)+
{1\over 4\pi}\int d^2 x\,d^4\theta\,(\delta \Sigma+\delta \bar\Sigma)\, (K_c(\lambda,\bar \lambda) - K_{tc}(\tilde \lambda, \bar {\tilde \lambda})) \,.}

Next, we would like to check the invariance of \actiontwotwoAc\ under K\"ahler transformations.  It is trivially invariant under
\eqn\CKinv{\CK \to \CK + G(\tilde \lambda)+ \bar G(\bar {\tilde \lambda}) ~.}
In addition,
under the K\"ahler transformation
\eqn\CKinvc{\CK \to \CK + F( \lambda)+ \bar F(\bar { \lambda}) ~}
the anomaly
shifts  by the super-Weyl variation  \superweylA\ of the supersymmetric local counterterm \counterterm.

The lack of strict K\"ahler invariance under \CKinvc\ can be interpreted in several different ways with interesting consequences.  First, we can simply state that the K\"ahler transformation should be accompanied by a change in a local counterterm.  Second, we can assign a transformation law  to $\Sigma$
\eqn\SigmaKahler{ \Sigma \to \Sigma+ {6\over c} F(\lambda)}
and use the first term in \actiontwotwoAc\ to achieve full K\"ahler invariance of the anomaly (here we assume $c\neq 0$). This perspective means that $e^{{c\over 6}\Sigma}$ is a holomorphic section of a line bundle, whose first Chern class
is the cohomology class of the K\"ahler form on $\CM$ and therefore
$\CM$ must be Hodge.
This result, which we  have now derived using the anomaly, is known for sigma models with Calabi-Yau target spaces and for general $(2,2)$ theories.  It is also natural in the context of string compactification as a property of the four-dimensional supergravity theory \WittenHU\ (see a refinement of this statement in \refs{\KomargodskiPC\KomargodskiRB\SeibergQD-\BanksZN}).  In that context the action depends on $ -{c\over 6}(\Sigma + \bar \Sigma) +{\cal K}$ where $\Sigma$ is the spacetime dilaton superfield, or equivalently, it is the spacetime conformal compensator.
This is similar to our $\Sigma$, which is the two-dimensional conformal factor.
Indeed, integrating the anomaly \actiontwotwoAc\ we obtain the anomalous piece
of the effective action in superconformal gauge
\eqn\Seff{
\log Z_A\supset-{c\over 48\pi}\int d^2 x\, d^4\theta
\Big(\Sigma+\bar\Sigma-{6\over c}{\cal K}\Big)^2\,.}

Finally, we can try to use this analysis to suggest a stronger result.  It is well known, and we have used it extensively, that the anomaly variation is a well defined local term.  The lack of strict K\"ahler invariance means that our anomaly is not quite well defined. If the K\"ahler class of $\CM$ is trivial, there is no immediate problem  since we are not forced to perform the K\"ahler transformations~\CKinvc\ and we thus have a global description of the theory.  Different presentations of the theory might be related by K\"ahler transformations, but this can be absorbed in a local counterterm or in a redefinition of $\Sigma$.  However, when the K\"ahler class of $\CM$ is nontrivial, there is a difficulty. In that case we must cover $\CM$ with patches and transition functions that involve K\"ahler transformations and correspondingly a change in the  counterterm~\countertwo.

Now, consider the couplings changing in spacetime in such a way that we must use the transition functions (e.g.\ spacetime wraps a nontrivial cycle in $\CM$).  Here different parts of spacetime have coupling constants in different patches in $\CM$ and since the transition functions between them need a counterterm, e.g.\ \countertwo, there is no single Lagrangian in all of spacetime!  We suggest that such a situation is inconsistent.  This would mean that $\CM$ and the various fields on it are such that no such transition functions are needed.  Therefore, we arrive at the conclusion that $\CM$ must have vanishing K\"ahler class.  This argument is analogous to that of~\NekrasovWG,  with the difference being that we are considering the properties of the space of theories rather than the usual target space of a specific sigma model.

More generally, we should always require that the scale variation of the partition function is a local, globally-defined functional of the background fields. In our context, the anomaly functional contains $K$ explicitly and is therefore not invariant under K\"ahler transformations. The anomaly functional is well defined only if the K\"ahler class vanishes.

Let us now extract some useful physical information from our anomalies \actiontwotwoAc.
It suffices for our purposes to  evaluate the anomaly keeping only the bottom components of the multiplets of the exactly marginal parameters $\lambda^I$ and $\tilde\lambda^A$ and of $\delta \Sigma$
\eqn\Sigmasa{\delta \Sigma\bigr|=\delta \sigma + i\delta a ~,}
where $\delta a$ parameterizes the $U(1)_V$ transformation.  We find
\eqn\actiontwotwo{
\eqalign{
&\delta_{\Sigma}\log Z_A
=-{1\over 2\pi}\int d^2 x\,
\Bigg(\delta \s\, \left(g_{I\bar J }\partial_\mu\lambda^I\,\partial^\mu\bar\lambda{}^{\bar J}+\tilde g_{A\bar B}\,\partial^\mu\tilde\lambda^A\partial_\mu\bar {\tilde\lambda}{}^{\bar B}\right)\, -\half \square\delta \s K_c
\cr
&\qquad \qquad \qquad \qquad + \delta a
 	\left(\partial^\mu\cal{A}_\mu+\epsilon^{\mu\nu}\partial_\mu\tilde{\cal{A}}_\nu\right) +{c\over6}\,\Big( \delta \sigma\, \square\sigma+\delta a\,\square a\Big)
\Bigg)~, \cr
&g_{I\bar J }=\partial_I\partial_{\bar J}K_c ~,\cr
&\tilde g_{A\bar B}=\partial_A\partial_{\bar B} K_{tc}~,\cr
&{\cal A}_\mu= {i\over 2}\left(\partial_I K_c\partial_\mu\lambda^I-\partial_{\bar I}K_c \partial_\mu\bar\lambda^{\bar I}\right) ~,\cr
&\tilde{\cal A}_\mu = {i\over 2} \left(\partial_A K_{tc}\partial_\mu\tilde\lambda^A-\partial_{\bar A}K_{tc}\partial_\mu \bar{\tilde\lambda}{}^{\bar A}\right)~.
}}
Here we have integrated by parts and used the metrics $g_{I\bar J }$ and $\tilde g_{A\bar B}$  on $\CM_c$ and $\CM_{tc}$ as well as the pull-back of the K\"ahler one-forms ${\cal A}_\mu$ and $\tilde{\cal A}_\mu $, whose exterior derivatives give the K\"ahler two-forms of $K_c$ and $K_{tc}$.
We note that only the K\"ahler potential $K_c$ for the chiral multiplets appears in the term proportional to $\square \delta \s$ in \actiontwotwo. This is due to the fact that $\delta \Sigma$ is a chiral multiplet.  This point will be important below.

Another way of stating our equations uses the supercurrent multiplet.  As we review in Appendix C, the relevant axial supercurrent multiplet consists of real (in Lorentzian signature) $\CJ_{\pm\pm}$ and a chiral $\CW$ satisfying
\eqn\Amula{\bar D_{\pm}  \CJ_{\mp\mp}=\pm D_\pm\CW ~. }
In a conformal theory $\CW=0$.  Our anomaly is
\eqn\anomaoptt{\CW = -{c\over 24\,\pi} \CR + {1\over 4\pi}\bar D^2 (K_c( \lambda, \bar  \lambda)-K_{tc}(\tilde \lambda, \bar {\tilde \lambda}))~,}
where $\CR=\bar D^2\bar \Sigma$ is the chiral curvature superfield. As above, it is invariant under K\"ahler transformations of $K_{tc}$, but not under K\"ahler transformations of $K_c$. We can absorb K\"ahler transformations of $K_c$ by improvements of the energy-momentum tensor multiplet.
Alternatively, we can make it invariant by also shifting $\Sigma$. From the first point of view it thus follows that if the K\"ahler class of $\CM$ is non-vanishing, upon letting the coupling constants wrap some two-cycle in $\CM$, we would not be able to define a single energy-momentum tensor throughout our two-dimensional space. This again suggests that the K\"ahler class of $K_c$ vanishes.

In the absence of supersymmetry, the last term $\square\delta\s K_c$ in the first line of \actiontwotwo\ would be cohomologically trivial and could be tuned away by an appropriate choice of regularization scheme. Indeed, this term is proportional to the variation of the local term, $\delta_\sigma \int d^2x \sqrt{\gamma}RK _c$. However, since we are, by assumption, defining the partition function using a supersymmetric regulator, cohomologically trivial terms must arise from the Weyl variation of $U(1)_A$ supergravity invariants. The most general such term is $\delta_\sigma \int d^2x \sqrt{\gamma}R(F(\lambda)+\bar F(\bar \lambda))$ with holomorphic $F(\lambda)$ (see \counterterm).  Therefore, modulo K\"ahler transformations, the anomaly in \actiontwotwo\ is a genuine new contribution to the trace anomaly in $\CN=(2,2)$ SCFTs.

Even in the absence of supersymmetry, the terms in~\actiontwotwo\
\eqn\zamolo{-{1\over 2\pi}\int d^2 x\delta \s\,\Big(  g_{I\bar J }\partial_\mu\lambda^I\,\partial^\mu\bar\lambda{}^{\bar J}+
  \tilde g_{A\bar B}\,\partial^\mu\tilde\lambda^A\partial_\mu\tilde\lambda^B\Big)~}
are cohomology nontrivial since they cannot be generated by the Weyl variation of any local term. These are precisely the terms we discussed in \OsEq. This  part of the anomaly is captured by a nonlocal term in the effective action, whose Weyl variation reproduces the anomaly \zamolo. Supersymmetry relates this nonlocal universal term to local terms that upon a Weyl transformation give rise to the term $\square \delta\s K_c$ in \actiontwotwo.
We can thus reconstruct these terms in the effective action by integrating the anomaly sigma model.

In the evaluation of the partition function for constant sources $\lambda^I$ and $\tilde \lambda^A$ and vanishing $a$,
it suffices to focus on the term $\square \delta \s K_c$.  First we covariantize it
\eqn\integra{
  \delta_{\Sigma}\log Z_A\supset{1\over 4\pi}\int d^2 x\sqrt{\gamma}\,\square\delta\s K_c~ ~.}
Using $\delta_\s \sqrt{\gamma} R=-2\sqrt{\gamma}\,\square\,\delta\s$ we learn that the partition function contains
\eqn\partyf{
  \log Z_A\supset -{1\over 8\pi}\int d^2x\sqrt\gamma RK_c~.}
We repeat that this is not a supersymmetric local term.  It is related by supersymmetry to some nonlocal terms that generate the anomaly~\zamolo. This is the reason the coefficient of~\partyf\ is physical.  The super-Weyl invariant terms  in $Z_A$ vanish identically  since all two-dimensional supergravity backgrounds are superconformally flat.

Upon evaluating the partition function on the two-sphere  $S^2$  and for constant sources we obtain that the $S^2$ partition function of
a $\CN=(2,2)$ SCFT regularized preserving $U(1)_A$ is
 \eqn\UoneVfinal{
 Z_A[{S^2}]=
\left({r\over r_0}\right)^{c\over 3}e^{-{1\over 8\pi}\int_{S^2} d^2x\sqrt\gamma RK_c }=\left({r\over r_0}\right)^{c\over 3}e^{-K_c}~,}
where we exhibit the radius of the sphere $r$, which arises from the ordinary central charge anomaly.  Note that in agreement with the picture above, we either say that $Z_A$ is not K\"ahler invariant, or we accompany K\"ahler transformations with $r \to r e^{{3\over c}(F(\lambda) + \bar F(\bar \lambda))}$. We remark that \UoneVfinal\ is correct
for any compact manifold with the topology of the two-sphere, the
prefactor being reexpressed in terms of the area of the manifold. This is consistent with~\GomisWY, who argued that the $S^2$ partition function is independent of squashing.

The analysis extends almost verbatim in $U(1)_V$ supergravity. In this case the anomaly \actiontwotwoAc\ becomes
\eqn\anomVcal{
\delta_{\tilde\Sigma}\log Z_V
\supset{1\over 4\pi}\int d^2 x\,d^4\theta\, (\delta \tilde \Sigma+\delta  \bar{\tilde \Sigma})\left(K_{tc}(\lambda,\bar\lambda)- K_{c}(\tilde\lambda,\bar{\tilde\lambda})\right)~.}
And using $\tilde \Sigma = \sigma + i\tilde a$
\eqn\actiontwotwoV{
\eqalign{
\delta_{\tilde\Sigma}\log Z_V
=-{1\over 2\pi}\int d^2 x\,
&\Bigg(\delta \s\, \left(g_{I\bar J }\partial_\mu\lambda^I\,\partial^\mu\bar\lambda{}^{\bar J}+\tilde g_{A\bar B}\,\partial^\mu\tilde\lambda^A\partial_\mu\bar {\tilde\lambda}{}^{\bar B}\right)\, -\half \square\delta \s K_{tc}
\cr
&+\delta \tilde a\left(\partial^\mu \tilde{{\cal A}}_\mu + \epsilon^{\mu\nu}\partial_\mu{\cal A}_\nu\right)+{c\over6}\Big(\delta\sigma\,\square\sigma+
\delta\tilde a\,\square\tilde a\Big)
\Bigg)~.
}}
Now the K\"ahler potential for the twisted chiral multiplets $ K_{tc}$ appears explicitly in the anomaly since $\tilde\Sigma$ is a twisted chiral multiplet. Integrating the anomaly, as above, we arrive at
 \eqn\UoneAinal{
 Z_V[{S^2}]=\left({r\over r_0}\right)^{c\over 3}e^{-{1\over 8\pi}\int_{S^2} d^2x\sqrt\gamma R K_{tc} }=\left({r\over r_0}\right)^{c\over 3}e^{- K_{tc}}~.}

In summary, we have re-derived  the result that supersymmetric $S^2$ partition functions of $\CN=(2,2)$ SCFTs
are expressed in terms of the K\"ahler potential on the appropriate moduli space of
theories~\refs{\JockersDK,\GomisWY} (see also \refs{\GerchkovitzGTA}). Our derivation shows that this phenomenon follows directly from a new trace anomaly in supersymmetric field theories. The new trace anomaly is tied by supersymmetry to the anomaly associated with the Zamolodchikov metric.  Our methods also led to the suggestion that the K\"ahler class of $\CM$ vanishes.

It is important that the anomalies we discussed reflect UV physics.  They are independent of the background spacetime and can be explored locally in flat space.  The sphere partition function was used as a tool to extract this anomaly.  We note that we simply substituted $\Sigma$ of a sphere.

In Appendix D we discuss a classification of supersymmetric backgrounds using our superconformal gauge formalism.  Specifically, we consider $\Sigma = \sigma + ia + \theta^2w$ with various $\sigma$, $a$, and $w$. The round sphere discussed above corresponds to $a=0$ but with $\sigma$ and $w$ non-vanishing. We would like now to make a few comments on the other possible supersymmetric backgrounds, and in particular, about the topologically twisted background on the two-sphere $a=i\sigma$ (such that $\Sigma=\bar\CR=0$). Due to the anomaly ${c/6}\int d^2x \delta a \square a $ in~\actiontwotwo, the partition function needs to transform with a nonzero phase under $U(1)_V$ transformations. Hence, the $S^2$ partition function on the twisted sphere vanishes (see~\refs{\WittenXJ\WittenZZ-\BershadskyCX}). If one introduces the parameter $\bar w$, then this phase can be absorbed by including $\bar w^{-c/3}$  in the partition function. The partition function can be argued to be  holomorphic as a function of the coupling constants. But since the holomorphic counterterm~\counterterm\ does not vanish  (the anti-holomorphic one vanishes), only the singular part of the dependence on coupling constants is physical. It would be interesting to understand what these singular pieces mean.  They were recently computed in~\refs{\BeniniNOA,\ClossetRNA}. Another interesting open question concerns Calabi's diastasis, which is a nice K\"ahler invariant observable in $(2,2)$ SCFTs.  It has an elegant interpretation in terms of conformal interfaces~\BachasNXA, and it would be interesting to see if our methods shed light on it.

\newsec{(0,2) Supersymmetric Theories}

Here we consider $(0,2)$ SCFTs and study their trace anomaly.  Our conventions are such that the supersymmetry is right-moving. The exactly marginal operators are necessarily in Fermi multiplets~\Plesser\ (see~\WittenYC\ for background on $(0,2)$ models). The corresponding couplings are in chiral superfields $\lambda^I$. We will determine
their contribution to the  conformal anomaly.

We plan to place the theory in a nontrivial supergravity background.  This is simpler when the supergravity theory is anomaly free.  First, to avoid gravitational anomalies we must relate the left-moving and the right-moving central charges $c_L= c_R$.  Similarly, $c_R$ determines the anomaly in the right-moving $U(1)$ current and we assume that there is also a left-moving $U(1)$ current with the same anomaly.  Then we can gauge an anomaly free linear combination of these two currents.  Note that even if we do not have such an anomaly free setup, we can imagine adding decoupled fields to achieve it.

Under these assumptions, the supergravity transformations (which include gauge transformations for a $U(1)$ gauge field) are non-anomalous. One can then naturally couple the theory to the corresponding supergravity and study it in nontrivial supergravity backgrounds.  We will refer to the gauged $U(1)$ symmetry as axial (as in our discussion of $U(1)_A$ $(2,2)$ supergravity above) and then the vector $U(1)$ symmetry is a global symmetry, suffering from an anomaly.

Before delving into a technical discussion, let us summarize what we find. The trace anomaly in $(0,2)$ models contains a term depending on the K\"ahler potential, supersymmetrizing the ordinary bosonic anomaly~\OsEq. But there is an additional term in the trace anomaly that depends on a new function of the couplings, $H$. This function is not fixed by the $(0,2)$ theory. It depends on precisely how we couple the theory to the background fields.  It is instructive to consider a $(2,2)$ theory viewed as a $(0,2)$ theory.  Then this function $H$ is given as
\eqn\zerotwointint{H\sim K_c-K_{tc}~;}
i.e.\ in this case the function is physical and unambiguous. But in general $(0,2)$ models it is ambiguous. The sphere partition function depends on $H$ and therefore the sphere partition function in such theories is not universal.
However, the arguments leading to the conclusion that the K\"ahler class vanishes do hold in $(0,2)$ models and in particular, the moduli space of SCFTs cannot be compact. This is in accord with intuition from the heterotic string, where $(0,2)$ models are used to construct $\CN=1$ supergravity theories in spacetime. In such cases it is known that the vacuum manifold is K\"ahler-Hodge~\WittenHU\ (see also~\refs{\KomargodskiPC\KomargodskiRB\SeibergQD-\BanksZN}).  For a related stringy discussion see~\refs{\PeriwalMX,\Plesser}.

As in the $(2,2)$ theory, we find it convenient to use the superconformal gauge.  But unlike the $(2,2)$ theory there are two natural ``conformal gauges.''  The difference between them is in the gauge condition imposed on the $U(1)$ gauge field.

One possibility is to use the gauge $A_{--}=0$, where $A_{--}$ couples to the right-moving $U(1)$ current $j_{++}$ in the superconformal algebra.  In this case the remaining degrees of freedom are in a real superfield
 $V=\sigma + {i\over 2}\theta^+\Psi_{+} + {i\over 2} \bar \theta^+\bar \Psi_{+} + \theta^+\bar \theta^+ A_{++}$. In the linearized supergravity approximation which we review in Appendix C (see also~\DumitrescuIU) this corresponds to $\CH_{++}=\CH_{----}=0$ with $V={1\over 2}\CH$.

Alternatively, we impose Lorentz gauge on that gauge field $\partial^\mu A_\mu=0$, which is solved locally by $A_\mu =\epsilon_{\mu\nu}\partial^\nu a$.  Then the remaining degrees of freedom are in a chiral multiplet $\Sigma=\sigma+ ia + i\theta^+\Psi_{+} -i\theta^+\bar\theta^+\partial_{++} (\sigma+ia)$.

These two multiplets are almost identical.  Given $\Sigma$ we can write $V=\half(\Sigma +\bar \Sigma)$.  But given $V$, the chiral superfield $\Sigma = -{i \over \partial_{++}}\bar D_+ D_+ V$ is nonlocal.  The lack of locality affects only $a$.  Its zero mode is present in $\Sigma$ but not in $V$.  Conversely, a constant mode of $A_{++}$ correspond to a linearly growing $a$.

The gauge invariant chiral curvature superfield can be expressed using either of these fields
\eqn\chiralcur{\CR_- =2 \partial_{--} \bar D_+  V = \partial_{--} \bar D_+ \bar \Sigma~.}

Super-Weyl transformations are associated with a chiral $\delta \Sigma$ and simply shift $\Sigma$.  Their action on $V$ is $2\delta V = \delta \Sigma + \delta \bar \Sigma$.  The shift of $a$ by a constant represents the action of the global vector $U(1)$ symmetry that is not gauged.  This global symmetry shifts $\Sigma $ by an imaginary constant and does not act on $V$.

Then, the most general expression for the anomaly action is
\eqn\anozerotwo{\eqalign{
\delta_\Sigma \log Z =&{c\over 12\pi} \Big(i\int d^2x d\theta^+ \delta \Sigma \CR_- + c.c.\Big)\cr
&+ {i\over 4\pi}\int d^2x d\theta^+d\bar \theta^+
\Big((\delta\Sigma+\delta \bar \Sigma) A_I \del_{--} \lambda^I+(\delta \Sigma-\delta \bar \Sigma) B_I \del_{--} \lambda^I\Big)+\hbox{c.c.} ~.}}
So far, $A_I,B_I$ are arbitrary functions of the couplings.

The first term is the ordinary central charge anomaly.  It can also be written as follows in our two slightly different versions of conformal gauge: \eqn\confa{i{c\over 12\pi} \int d^2 x d^2\theta^+ (\delta \Sigma -\delta \bar \Sigma) \partial_{--} (\Sigma+ \bar \Sigma) = i{c\over 6\pi} \int d^2 x d^2\theta^+ (\delta \Sigma -\delta \bar \Sigma) \partial_{--} V~.}
The functions $A_I,B_I$ represent the anomalies that arise in the presence of exactly marginal coupling constants. Already from this we can infer that the metric on $\CM$ is Hermitian.

The components expansion of \anozerotwo\ leads to a term proportional to $\delta \sigma\epsilon^{\mu\nu}(\del_{\bar J}A_I-\del_IA_{\bar J})\del_\nu\lambda^I\del_\mu\bar\lambda^{\bar J}+c.c.$. However as we discussed in section 2, this is consistent only if (locally) $A_I=\del_I K$. Furthermore, $K$ has to be real in order to eliminate type-A anomalies that are present upon expanding the first term.
Similar considerations show that $B_I=\del_I H$ with some real function $H$.

Therefore, the anomaly must be of the form
\eqn\anozerotwois{ \eqalign{
\delta_\Sigma \log Z =  {i\over 4\pi}\int d^2x d^2\theta^+&\Bigg(  (\delta \Sigma - \delta \bar \Sigma) \partial_{--} \left({2c\over 3} V +H\right) \cr
&+(\delta \Sigma+\delta \bar \Sigma)\left( \del_IK \del_{--} \lambda^I-\del_{\bar I}K \del_{--} \bar\lambda^{\bar I}\right)\Bigg) ~.}}

The expression \anozerotwo\ satisfies the usual consistency conditions including the Wess-Zumino conditions.  We therefore see that our additional considerations in section 2 concerning which anomalies are allowed show that the metric on $\CM$ must be K\"ahler (in accord with intuition from heterotic compactifications, which lead to $\CN=1$ supergravities).  Below we will also find some global restrictions on $\CM$.

As in \Amula-\anomaoptt, we can express the anomaly as an operator statement.  The theories we study have a supercurrent multiplet with real $R_{\pm\pm}$ and $\CT_{----}$ satisfying \DumitrescuIU\ (see Appendix C)
\eqn\Rmulzerotwof{\eqalign{&\del_{--}R_{++}+\del_{++}R_{--}=0~,\cr& \bar D_+\left(\CT_{----} - i \del_{--}R_{--}\right)=0~.}}
When the theory is conformal we also have $\bar D_+R_{--}=0$.  $R_{--}$ is  the left-moving current that we assumed exists in the CFT.  Our anomaly modifies $\bar D_+R_{--}=0$ to
\eqn\anomaopzt{\bar D_+R_{--}  = i{c\over 12\pi} \CR_-
+{i\over 4\pi}\bar D_+(\partial_I K \partial_{--} \lambda^I- \partial_{\bar I} K \partial_{--} \bar \lambda^{\bar I}
+\partial_{--} H  )~.}

Next we should identify the ambiguity (i.e.\ the cohomologically trivial terms that arise from variations of $(0,2)$-supersymmetric local counterterms). This will allow us to determine the actual  anomaly.  For that we should supersymmetrize \countertwo.  We can either use the full nonlinear supergravity (see e.g.\ \NibbelinkWB), or simply use linearized supergravity as in \DumitrescuIU\ and Appendix C to show that the local counterterm is
\eqn\coutzerotwo{
-{i\over 8\pi}\int d^2 x d\theta^+ h(\lambda^I)\CR_- ~, }
where $h(\lambda^I)$ is holomorphic. This counterterm allows us to absorb some holomorphic transformations on $H$ and $K$ but most of the information in \anozerotwois\ is cohomologically nontrivial.

Even though the anomaly associated with $H$ seems like a nontrivial anomaly, which cannot be absorbed in a local counterterm, in fact, it is not physical in $(0,2)$ theories.  It can be absorbed in a redefinition of $V$.\foot{Note that we cannot absorb $H$ in $\Sigma+\bar \Sigma$ by a local transformation. However, this redefinition is indeed a truly local transformation in $(0,2)$ supergravity. It can be understood in linearized supergravity before picking any gauge. There we simply shift $\CH$ by the function $H(\lambda^I,\bar\lambda^{\bar I})$. This modifies the couplings of the theory to curved space by additional terms in the Lagrangian, which depend on the coupling constants. For conformal theories, this modification only depends on derivatives of the coupling constants. Such ambiguities do not play a role in $(2,2)$ theories.}  Physically, this means that we redefine the metric and its superpartners by some function of the coupling constants $\lambda^I,\bar \lambda^{\bar I}$. In other words, when we allow the couplings to be general functions, we can add new terms to the Lagrangian that vanish upon setting the couplings to constants. Such a freedom exists in $(0,2)$ theories and it leads to the anomaly
$H$. There is no a priori principle that fixes $H$, unless the theory is a $(2,2)$ theory in which case this freedom does not exist and $H$ becomes physical~\zerotwointint.

After removing $H$ we conclude that the anomaly can be written as
\eqn\anozerotwoi{
\delta_\Sigma \log Z =
i{c\over 12\pi}\left(\int d^2x d\theta^+\delta\Sigma{\cal R}_- -c.c.\right)
+ {i\over 4\pi}\int d^2x d^2\theta^+  (\delta \Sigma + \delta \bar \Sigma) \Big(\del_IK \del_{--} \lambda^I -c.c.\Big) ~.}

Expanding it in components with the only nonzero background fields   $\Sigma = \s + i a $ and  $\lambda^I|$ we find
\eqn\evali{\eqalign{
&\delta_\Sigma \log Z = -{1\over 2\pi}\int d^2x
\Big({c\over 6} (\delta \sigma\square \sigma + \delta a \square a)\cr
&\qquad\qquad \qquad\qquad  + \delta\sigma G_{I\bar J} \del^\mu \lambda^I\del_\mu \bar\lambda^{\bar J}-{1\over 4}\square\delta \sigma K
+{1\over 4}\delta a\left(
	\partial^\mu{\cal A}_\mu
	+\epsilon^{\mu\nu}\partial_\mu{\cal A}_\nu
 \right)
 \Big)~, \cr
 &G_{I\bar J}=\partial_I\partial_{\bar J} K~, \cr
 &{\cal A}_\mu= i\left(\del_IK \del_\mu \lambda^I-\del_{\bar I}K\del_\mu \bar\lambda^{\bar I}\right) ~.}}
The first term is the ordinary anomaly.  The second term is the anomaly \OsEq\ associated with the metric on $\CM$.

The situation with K\"ahler transformations is as in $(2,2)$ theories.  K\"ahler transformations  can be absorbed in a shift of $\Sigma$.  As there, $e^{{c\over 6}\Sigma}$ is a holomorphic section of a line bundle over $\CM$ and therefore $\CM$ is not only K\"ahler, but it is also Hodge.  Also, as in $(2,2)$ theories, we suggest that the K\"ahler class of $\CM$ is in fact trivial.

These results are consistent with the expectation from the string application of these models.  When the $(0,2)$ theory is used as the worldsheet of a compactified heterotic string it leads to $\CN=1$ supersymmetry in four dimensions and $\CM$ is the target space of some of its chiral superfields.  In this case it is known that $\CM$ should be K\"ahler~\refs{\PeriwalMX,\Plesser} and Hodge~\WittenHU.  We extend these conclusions to all $(0,2)$ SCFTs and argue that the K\"ahler class of $\CM$ should be trivial.

While the anomaly functional~\evali\ contains the term $\square\delta\sigma K$, the partition function depends on the choice of $H$ (which we have set to zero for simplicity) and therefore it is not universal.

\newsec{$\CN=2$ Supersymmetry in $d=4$}

We now proceed to the supersymmetric generalization of the conformal anomaly~\OsEqfourd. For $\CN=2$ supersymmetry the appropriate superspace expression is
\eqn\dZmoduli{
\delta_\Sigma\log Z\supset{1\over 192\pi^2}\int d^4 x\,d^4\theta\,d^4\bar\theta
\,E(\delta\Sigma+\delta\bar\Sigma)K(\lambda^I,\bar\lambda^{\bar I})\,.}
The super-Weyl parameters $\delta\Sigma$ and $\delta\bar\Sigma$ are chiral and anti-chiral
superfields, respectively. They can
be viewed as a conformal compensator in $\CN=2$ supergravity~\deWitTN.
$\lambda^I$ and $\bar\lambda^{\bar I}$ are chiral and anti-chiral superfields
with Weyl weight zero, whose lowest components are the exactly marginal couplings,
which we also denote as $\lambda^I$ and $\bar\lambda^{\bar I}$.
$K(\lambda,\bar\lambda)$ is the K\"ahler potential on the conformal manifold $\CM$.

In addition to the anomaly that contains the moduli, we also have the
usual Weyl anomaly, which depends only on the supergravity multiplet.
Its superspace expression is an integral over chiral $\CN=2$ superspace (see e.g.\ \KuzenkoGVA\ and Appendix B for details)
\eqn\dZSUGRA{\eqalign{
\delta_\Sigma\log Z&\supset {1\over16\pi^2}\int d^4 x\,d^4\theta\,{\cal E}\delta\Sigma
\Big(a\,\Xi+(c-a)\,W^{\alpha\beta}W_{\alpha\beta}\Big)+\hbox{c.c.}\cr
&\supset{1\over16\pi^2}\int d^4 x\,\sqrt{\gamma}\,\delta\sigma
\left(c\,C^{\mu\nu\rho\s}C_{\mu\nu\rho\s}
-a\Big(E_4-{2\over3}\square R\Big)\right)~,
}}
where $W^{\a\b}$ and $\Xi$ are chiral superfields.
$W^{\a\b}$ is the Weyl superfield, while $\Xi$ is constructed from curvature
superfields that appear in the commutators of super-covariant
derivatives in curved superspace.

To work out the component field expansion of \dZmoduli, we need to
know the component expansion of an action of the general form
\eqn\action{
S={1\over 4}\int d^4 x\, d^4\theta\, d^4\bar\theta\,
E\,{\cal K}(\lambda^A,\bar\lambda{}^{\bar A})~,}
where $\lambda^A$ and $\bar\lambda{}^{\bar A}$ are ${\cal N}=2$ chiral and anti-chiral
multiplets with Weyl weight $w=0$, respectively. For our anomaly \dZSUGRA\ we will then specify to
\eqn\Kaehler{
{\cal K}(\lambda^A,\bar\lambda{}^{\bar A})=
{1\over192\pi^2}(\delta\Sigma+\delta\bar\Sigma)\,K(\lambda^I,\bar\lambda{}^{\bar I})\,.}

For calculating the component expansion of \action\ we follow \deWitZA.
We start with the special case ${\cal K}={\cal A}\,\bar{\cal B}$, where
${\cal A}$ and $\bar {\cal B}$ are chiral and anti-chiral multiplets
respectively. Keeping only the bottom components ${\cal A}|=A$, $\bar{\cal B}|=\bar B$,
and the metric background (i.e. dropping the bosonic auxiliary fields in the supergravity multiplet), we get
\eqn\dWfourtwo{
S=\int d^4 x\sqrt{\gamma}\left(\nabla^2 A\,\nabla^2\bar B-2\,\nabla_\mu A
\left(R^{\mu\nu}-{1\over3}R\,\gamma^{\mu\nu}\right)\nabla_\nu\bar B\right)~.}
In order to find the answer for a generic ${\cal K}(\lambda^A,\bar\lambda{}^{\bar A})$ we expand around a reference point and then use the fact that product of chiral multiplets with Weyl weight $w=0$ is a chiral multiplet with $w=0$ and similarly for anti-chiral multiplets.  Then ${\cal K}(\lambda^A,\bar\lambda{}^{\bar A})$ can be expressed as a sum $\sum_i \CA_i \bar \CB_i$ with chiral $\CA_i$ and anti-chiral $\bar \CB_i$ and we can use \dWfourtwo\ for each term in the sum.
Doing this we arrive at
\eqn\LL{\eqalign{
S&=\int d^4 x\sqrt{\gamma}\Big\lbrace{\cal K}_{AB\bar C\bar D}\nabla^\mu\lambda^A\,
\nabla_\mu\lambda^B\,\
\nabla^\nu\bar\lambda{}^{\bar C}\,\nabla_\nu\bar\lambda{}^{\bar D}
+{\cal K}_{AB\bar C}\nabla^\mu\lambda^A\,\nabla_\mu\lambda^B\,\square\bar\lambda{}^{\bar C}\cr
\noalign{\vskip.3cm}
&\qquad+{\cal K}_{\bar A\bar B C}\nabla^\mu\bar\lambda{}^{\bar A}\,
\nabla_\mu\bar\lambda{}^{\bar B}\,\square\lambda^C
+{\cal K}_{A\bar B}\,\square\lambda^A\,\square\bar\lambda{}^{\bar B}\cr
\noalign{\vskip.2cm}
&\quad\qquad-2\,{\cal K}_{A\bar B}\nabla_\mu\lambda^A\Big(R^{\mu\nu}
-{1\over3}R\,\gamma^{\mu\nu}\Big)\nabla_\nu\bar\lambda{}^{\bar B}\Big\rbrace
~,}}
where $\lambda^A=(\Sigma,\lambda^I$) ($\bar\lambda{}^{\bar A}
=(\bar\Sigma,\bar\lambda{}^{\bar I})$.
Using \Kaehler\ and the following definitions for the metric, connection and curvature
on a K\"ahler manifold with K\"ahler potential ${K}$
\eqn\RiemannKaehler{\eqalign{
&{g}_{I\bar J}=\p_I\p_{\bar J}{K}~,\cr
\noalign{\vskip.2cm}
&\Gamma^{I}_{JK}={g}^{I\bar L }\p_J\p_K\p_{\bar L}{K}~,\cr
&{\cal R}_{I\bar J K\bar L}
=\p_I\p_{\bar J}{g}_{K\bar L}
-{g}^{M\bar N}\,\p_{I}{g}_{K\bar L}\,
\p_{\bar J}{g}_{M\bar N}~,}}
we arrive, after several integrations by parts, at
\eqn\actiontwo{\eqalign{
&\delta_\Sigma\log Z\supset{1\over96 \pi^2}\int d^4 x\sqrt{\gamma}\Bigg\lbrace
\delta\sigma{\cal R}_{I\bar K J\bar L}\nabla^\mu\lambda^I\,
\nabla_\mu\lambda^J\,\nabla^\nu\bar\lambda{}^{\bar K}\,\nabla_\nu\bar\lambda{}^{\bar L}\cr
\noalign{\vskip.2cm}
&+\delta\sigma{g}_{I\bar J}\left(\hat{\square}\lambda^I\,\hat{\square}\bar\lambda{}^{\bar J}
-2\Big(R^{\mu\nu}-{1\over3}R\,\gamma^{\mu\nu}\Big)\nabla_\mu\lambda^I\,
\nabla_\nu\bar\lambda{}^{\bar J}\right)\cr
\noalign{\vskip.2cm}
&+{1\over 2}\,K\,\square^2\delta\sigma+{1\over6}K\,\nabla^\mu R\,\nabla_\mu\delta\sigma
+K\,\left(R^{\mu\nu}-{1\over3}\gamma^{\mu\nu}R\right)\nabla_\mu\nabla_\nu\delta\sigma\cr
\noalign{\vskip.2cm}
&-2\,{g}_{I\bar J}\,\nabla^\mu\lambda^I\,\nabla^\nu\bar\lambda{}^{\bar J}\,
\nabla_\mu\nabla_\nu\delta\sigma
+i\,{g}_{I\bar J}\left(\hat\nabla^\mu\hat\nabla^\nu\lambda^I\,\nabla_\nu\bar\lambda{}^{\bar J}
-\hat\nabla^\mu\hat\nabla^\nu\bar\lambda{}^{\bar J}\,
\nabla_\nu\lambda^I\right)\nabla_\mu\delta a\cr
\noalign{\vskip.2cm}
&-{i\over2}\left(\hat\nabla_I\hat\nabla_J K\,\nabla^\mu\lambda^I\nabla_\mu\lambda^J
-\hat\nabla_{\bar I}\hat\nabla_{\bar J}K\,
\nabla^\mu\bar\lambda{}^{\bar I}\nabla_\mu\bar\lambda{}^{\bar J}
+\nabla_I K\,\hat{\square}\lambda^I
-\nabla_{\bar I}K\,\hat{\square}\bar\lambda{}^{\bar I}\right)
\square\delta a\cr
\noalign{\vskip.2cm}
&+i\left(R^{\mu\nu}-{1\over 3}R\,\gamma^{\mu\nu}\right)\left(\nabla_I K\,\nabla_\mu\lambda^I
-\nabla_{\bar I}K\,\nabla_\mu\bar\lambda{}^{\bar I}\right)\nabla_\nu\delta a
\Bigg\rbrace ~.
}}
As in~\OsEqfourd, the hats denote covariant derivatives with respect to
target space diffeomorphisms
acting on the $\lambda^I$. Note that this action is completely covariant under
target space diffeomorphisms and we nicely identify the term~\OsEqfourd\
in the second line
of~\actiontwo. We also identify the new anomaly~\addano\ in the first line.
It appears with the
Riemann tensor of $\CM$. To take into account the complete anomaly we
have to add \dZSUGRA\ to \actiontwo.

Using the expressions in Appendix B one realizes that the terms in the third
line of~\actiontwo\  can be written as a variation of a local term. Specifically,
\eqn\cohotrivial{\eqalign{
&\sqrt{\gamma}\left({1\over 2}\square^2\delta\s+{1\over6}\nabla^\mu R\,\nabla_\mu\delta\s
+\left(R^{\mu\nu}-{1\over3}\gamma^{\mu\nu}R\right)\nabla_\mu\nabla_\nu\delta\s\right)
=\sqrt{\gamma}\,{1\over2}\,\Delta_4\,\delta\s\cr
&\qquad\qquad\qquad=
\delta_{\s}\left(\sqrt{\gamma}\left[{1\over8}E_4-{1\over12}\square R+c\,C^2\right]\right)\,,}}
where $c$ is an arbitrary function of the moduli and
$\Delta_4$ the FTPR operator (see Appendix B).
For $c=0$ this is precisely the combination that appears in the
$\CN=2$ supersymmetric version of the Gauss-Bonnet invariant (see e.g.\ \ButterLTA).
The supersymmetric Gauss-Bonnet term may include as a prefactor an arbitrary holomorphic
function of the moduli. Therefore,
\actiontwo\ is cohomologically trivial if $K=F+\bar F$
is a sum of a holomorphic and anti-holomorphic function of the moduli, in which
case it reduces to
\eqn\actiontwohol{
{1\over 192\pi^2}\int d^4 x\,\sqrt{\gamma}\left(F\Delta_4\bar\s+\hbox{c.c.}\right)\,.}
Indeed, consider the following local superspace counterterm
\eqn\GBKuzenko{
\int d^4 x\, d^4\theta\, {\cal E} F(\lambda)\,\big(\Xi - W^{\a\b}W_{\a\b}\big)+{\rm c.c.}}
The combination $\Xi-W^{\a\b}W_{\a\b}$ contains the Euler combination
$E_4-{2\over3}\square{R}$ and its Weyl variation is the supersymmetrization
of \actiontwohol.

To arrive at the $S^4$ partition function we simply need to integrate the combination
appearing in~\cohotrivial\ on $S^4$. Using
\eqn\intS{
\int_{S^4}\sqrt{\gamma}\,\left(E_4-{2\over3}\square R+c\,C^2\right)
=64\,\pi^2~}
we find\foot{Since the $S^4$ background is superconformally flat, the super-Weyl invariant terms in $Z$ vanish.}
\eqn\expectation{
Z[S^4]=\left({r\over r_0}\right)^{-4a}\,e^{K/12}~,}
as claimed in~\GerchkovitzGTA\ and~\GomisWOA.
We note that \expectation\
is true for any superconformally flat compact four-manifold if we express the
prefactor in terms of its volume.

The K\"ahler ambiguity $K\to K+{F}+\bar{F}$ of the partition function is taken care of by the ambiguous local counterterm~\GBKuzenko. As in $d=2$,  we will now see that
the trace anomaly is invariant under a correlated K\"ahler shift and Weyl transformation.

To find the change of the anomaly polynomial \dZSUGRA\ under an
infinitesimal  Weyl transformation
$\delta\tilde\Sigma$, we use (cf. \KuzenkoGVA)
\eqn\deltaXi{\delta_{\tilde\Sigma}\Xi=2\,\delta\tilde\Sigma\,\Xi
-2\,\bar\Delta\,\delta\bar{\tilde\Sigma}~,}
where $\bar\Delta$, the chiral projection operator,
is the $\CN=2$ supersymmetric generalization of the FTPR operator
(see also Appendix B). The Weyl superfield $W^{\a\b}$ transforms homogeneously with
weight one, while the chiral superspace density ${\cal E}$ transforms with weight $-2$
(the full superspace density $E$ is invariant).
We then find
\eqn\deltafivetwo{
\delta_{\tilde\Sigma}\delta_{\phantom{\tilde\Sigma}\!\!\!\!\Sigma}\log{Z}=
-{a\over 8\pi^2}\int d^4x\,d^4\theta\, {\cal E}\,\delta\Sigma\,\bar\Delta\,
\delta\bar{\tilde\Sigma}+\hbox{c.c.}\,.}
On the other hand, under a K\"ahler shift
$K\to K+F+\bar F$ it transforms as
\eqn\deltafiveone{
\delta_F\delta_\Sigma\log{Z}
={1\over192\pi^2}\int d^4 x\,d^8\theta\, E(\delta\Sigma\,\bar F
+\delta\bar\Sigma\,F)
={1\over 192\pi^2}\int d^4 x\,d^4\theta\,
{\cal E}\,\delta\Sigma\,\bar\Delta\ \bar F+\hbox{c.c.}}
Therefore, choosing
\eqn\deltaST{
\delta\tilde\Sigma={1\over 24 a}F~,}
the anomaly polynomial is invariant under an infinitesimal joint K\"ahler-Weyl transformation and therefore also under a finite transformation.
The invariance can be explicitly seen to hold for the partition
function \expectation.

As in two dimensions, this means that $\CM$ is not only K\"ahler, but it is also Hodge.
In addition, using  background $\lambda^I$ that vary in spacetime and wrap a
nontrivial cycle in $\CM$, we argue that the K\"ahler class of $\CM$ must be trivial. (For certain cases with an $\CN=4$ AdS$_5$ dual, it has been argued in~\LouisDCA\ that $\CM$ is special-K\"ahler. It would be interesting to understand when this happens in general.)

\bigskip
\noindent {\bf Acknowledgments:}

We would like to thank C.~Bachas, C.~Closset, S.~Cremonesi, L. Di Pietro, N. Ishtiaque, S.~Kuzenko, D.~Morrison,   Y.~Nakayama, V.~Niarchos, H.~Ooguri, H.~Osborn, K.~Papadodimas, K.~Skenderis,  E.~Witten, and J.~Zhou    for useful discussions.  Z.K. would like to thank the Perimeter
Institute for its very kind hospitality during the course of this project. The work of NS was supported in part by DOE grant DE-SC0009988.  Z.K. is supported by the ERC STG grant 335182, by the Israel Science Foundation under grant 884/11, by the United States-Israel Bi- national Science Foundation (BSF) under grant 2010/629
as well as by the Israel Science Foundation center for excellence grant (grant no. 1989/14). A.S. and Z.K. are supported by the I-CORE Program of the Planning and Budgeting Committee.
P.H. is supported by Physics Department of Princeton University.
A.S. and S.T. acknowledge support from GIF -- the German-Israeli Foundation for
Scientific Research and Development.
This research was supported in part by
Perimeter Institute for Theoretical Physics. Research at Perimeter Institute is supported by the
Government of Canada through Industry Canada and by the Province of Ontario through the Ministry
of Research and Innovation. J.G. also acknowledges further support from an NSERC Discovery
Grant and from an ERA grant by the Province of Ontario.
Any opinions, findings, and conclusions or recommendations expressed in this material are those of the authors and do not necessarily reflect the views of the funding agencies.

\bigskip

 \vfill\eject

\appendix{A}{Normalization of the Anomaly}

The normalization of the anomaly \OsEq\ and \OsEqfourd\ is fixed as follows. We compute the change of the contact term in $d=2$ under constant
rescaling of the coordinates $x\to e^{\lambda}x$ by writing the two-point function as
\eqn\oneoverx{ {1\over |x|^4}={1\over 32}\square^2\left({\log^2(x^2\mu^2)}\right)~,}
whose anomalous Weyl variation is
\eqn\deltaanom{
\delta^{(\rm anom)}_\lambda{1\over |x|^4}={\lambda\pi\over2}\square\,\delta^{(2)}(x)~.}
This is the contact term that is reproduced by the anomaly functional~\OsEq.

In $d=4$ we write
\eqn\confspace{
{1\over |x|^8}=-{1\over 768}\square^3\left({\log(x^2\mu^2)\over x^2}\right)~,}
whose anomalous Weyl variation is
\eqn\deltaxeight{
\delta^{(\rm anom)}_\lambda{1\over|x|^8}={\lambda\pi^2\over96}\square^2\delta^{(4)}(x)~.} This is matched by the anomaly~\OsEqfourd.

\appendix{B}{The FTPR operator and its properties}

We collect some properties  of the Fradkin-Tseytlin-Paneitz-Riegert operator
\Fradkinetal. It arises in our context  in the case that there is
only one exactly marginal modulus and one integrates \OsEqfourd\
by parts. One obtains, up to cohomologically trivial terms, that the anomaly is
\eqn\eqone{
\int d^4 x\,\sqrt{\gamma}\,\delta \sigma\lambda\Delta_4\lambda\,,}
where
\eqn\eqtwo{
\Delta_4=\square^2+{1\over3}\nabla^\mu R\nabla_\mu
+2\, R^{\mu\nu}\nabla_\mu\nabla_\nu-{2\over3}R\,\square}
is the FTPR-operator with the defining property that under Weyl
rescaling of the metric,
\eqn\eqthree{
\Delta_4 \to e^{-4\delta\s}\Delta_4}
when it acts on a scalar.
Another property of $\Delta_4$ which is used in Section 5 is
\eqn\eqfour{
\delta_\s\left(E_4-{2\over 3}\square R\right)
=-4\,\delta\s\,\left(E_4-{2\over 3}\square R\right)
+4\,\Delta_4\delta\s~.}~
This can be derived using
\eqn\deltaR{\eqalign{
\delta_\s R&=-2\,\delta\s\, R-6\,\square\delta\s~,\cr
\delta_\s R_{\mu\nu}&=-2\,\nabla_\mu\nabla_\nu\delta\s-g_{\mu\nu}\square\delta\s~,\cr
\delta_\s \square R&=-4\,\delta\s\,\square\,R
-2\,R\,\square\delta\s-2\nabla^\mu R\,\nabla_\mu\delta\s-6\,\square^2\delta\s~,}}
and the expression for the Euler density which we normalize to
\eqn\eqfive{
E_4=C^{\mu\nu\rho\sigma}C_{\mu\nu\rho\sigma}-2\,R^{\mu\nu}R_{\mu\nu}+{2\over3}R^2
}
such that
\eqn\eqsix{
\int_{S^4}d^4x\,\sqrt{\gamma}\,E_4= 64\,\pi^2\,.}
Here $C_{\mu\nu\rho\s}$ is the Weyl tensor.

In two dimensions instead of $\Delta_4$ we have $\Delta_2\equiv \square$.  It satisfies
$\Delta_2\to e^{-2\delta \sigma}\Delta_2$ and
$\delta E_2=\delta R=-2\delta\s R-2\,\Delta_2\,\delta\s$ with
$\int_{S^2}d^2 x\sqrt{\gamma}E_2=8\pi$.

The conformally covariant operators have generalizations
in chiral superspace. In $d=2$, $\CN=(2,2)$
this is the chiral projection operator $\bar\nabla^2$ which transforms
as $\bar\nabla^2\to e^{\delta\Sigma}\bar\nabla^2$
under super-Weyl transformations, which are parameterized by a chiral superfield
$\delta\Sigma$ (with a similar transformation for the anti-chiral projector $\nabla^2$).

For $\CN=2$ in $d=4$ the analog of the FTPR operator is the chiral
projection operator $\bar\Delta$
with the infinitesimal transformation
$\delta_\Sigma\bar\Delta=2\,\delta\Sigma\,\bar\Delta$
under a super-Weyl transformation parameterized by a
chiral scalar superfield $\delta\Sigma$. Its precise definition in terms of
super-covariant derivatives and curvature superfields is reviewed in
\KuzenkoGVA, where one also finds references to the original literature.

\appendix{C}{Review (and conventions) of two-dimensional supersymmetry}

\subsec{$(2,2)$}

We will use the notation $x^{\pm\pm}$ for the coordinates, which makes it easier to compare with spinors.  In Euclidean signature $x^{++} $ is the complex conjugate of $x^{--}$
and in Lorentzian signature they are two real independent coordinates.
The Ricci scalar is given by $R\sqrt{\gamma}=-\half\square\log \gamma$ in the conformal gauge, where $\gamma=\det \gamma_{\mu\nu}$.

The supercovariant derivatives, which can be obtained from the four-dimensional
ones of Wess and Bagger by dimensional  reduction, are
\eqn\scdtwo{
D_{\pm}={\p\over\p\theta^\pm}-i\,\bar\theta^\pm\p_{\pm\pm}\,,\qquad
\bar D_\pm=-{\p\over\p\bar\theta^\pm}+i\,\theta^\pm\p_{\pm\pm},}
The algebra is
\eqn\algebra{\{D_+,\bar D_+\} =2i \del_{++}~,\qquad \{D_-,\bar D_-\} =2i \del_{--}~.}

Chiral superfields $\lambda$ and twisted chiral superfields $\tilde \lambda$ are defined by
\eqn\rings{\eqalign{
&\bar D_\pm\lambda=0\cr
&\bar D_+\tilde\lambda=D_-\tilde\lambda=0 ~.}}

There are two interesting energy-momentum supermultiplets in a $(2,2)$ theory with an R-symmetry~\DumitrescuIU.  They are related by mirror symmetry.
First, there is the $U(1)_V$ supermultiplet
\eqn\Vmul{\bar D_{\pm}  \CR_{\mp\mp}=\pm \chi_{\mp}~,\qquad \bar D_+ \chi_\pm = \bar D_-\chi_\pm=0~,\qquad D_+\chi_-=\bar D_-\bar\chi_+ ~. }
It immediately follows that the bottom component, $\CR_{\mp\mp}\bigr|\equiv j^V_{\mp\mp} $ is a conserved vector current
\eqn\consv{\del_{++}j^V_{--}+\del_{--}j^V_{++}=0~.}
Often, there exists a twisted chiral operator $\tilde T$ such that
\eqn\chiT{\chi_+=\bar D_+ \bar {\tilde T} \qquad , \qquad \chi_-=- \bar D_-  \tilde T }
and then the last two equations in \Vmul\ are automatically satisfied and the first becomes
\eqn\Vmulb{\bar D_{+}  \CR_{--}=- \bar D_-  {\tilde T} \qquad , \qquad \bar D_{-}  \CR_{++}=- \bar D_+ \bar {\tilde T}~. }

The $U(1)_A$ supermultiplet
is obtained formally by acting with a mirror symmetry transformation on~\Vmul
\eqn\Amul{\eqalign{&\bar D_{\pm}  \CJ_{\mp\mp}=\pm \CY_{\mp}~,\qquad  D_{\pm} \CY_{\pm}=0~,\qquad \bar D_
\pm\CY_{\mp}=0~,\cr&    D_+\CY_-+D_-\CY_+=0~. } }
In this case $j^A_{\mp\mp}\equiv \CJ_{\mp\mp}\bigr| $ is a conserved axial current
\eqn\consa{\del_{++}j^A_{--}-\del_{--}j^A_{++}=0~.}
Often, there is a chiral $\CW$ such that
\eqn\CWdef{\CY_\pm= D_\pm\CW}
and then \Amul\ is replaced by
\eqn\Amulr{\bar D_{\pm}  \CJ_{\mp\mp}=\pm D_\mp \CW~.  }

Now we discuss linearized coupling to supergravity  (see also the analysis of~\ClossetPDA.)  We start from the case of $U(1)_A$ supergravity.
The supergravity multiplet is $(\CH_{\pm\pm}, \Sigma)$, where $\CH_{\pm\pm}$ is real (in Lorentzian signature) and $\Sigma$ is chiral. The linearized coupling to matter then takes the form
\eqn\linearizedcoupling{\delta \CL=\int d^4\theta \sum_{\pm}\CH_{\pm\pm}\CJ_{\mp\mp}+\left(\int d^2\theta \Sigma\CW+c.c.\right) ~.}
This is invariant under the linearized transformations
\eqn\gauge{\eqalign{& \delta \CH_{\pm\pm} = D_\pm\bar L_\pm-\bar D_\pm L_{\pm} ~,\cr &
\delta \Sigma=-\bar D^2(D_+L_--D_-L_+)~,}}
where $L_{\pm}$ are arbitrary superfields.
Note that the first line of~\gauge\ is consistent with the reality of $\CH_{\pm\pm}$ and that the action~\linearizedcoupling\ is invariant under~\gauge\ by using the defining relations~\Vmul.

In the superconformal gauge $\CH_{\pm\pm}=0$ and the only degree of freedom is $\Sigma = \sigma + ia+...$.  Here $\sigma$ is the conformal factor in the metric and $a$ represents the $U(1)_A$ gauge field in Lorentz gauge $A_\mu =\epsilon_{\mu\nu}\partial^\nu a$.  Under the super-Weyl transformations $\Sigma$ is shifted by a chiral superfield $\delta \Sigma$.  The shift of $a$ by a constant is a $U(1)_V$ global symmetry transformation.

The situation in $U(1)_V$ supergravity is analogous. In that case the only mode in the superconformal gauge is the twisted chiral superfield $\tilde \Sigma$.

\subsec{$(0,2)$}

For $(0,2)$ theories with a conserved $R$-symmetry the supercurrent multiplet consists of three superfields, $R_{\pm\pm},{\cal T}_{----}$, which are real in Lorentzian signature.  They satisfy~\DumitrescuIU
\eqn\Rmulzerotwo{\eqalign{&\del_{--}R_{++}+\del_{++}R_{--}=0~,\cr& \bar D_+\left(\CT_{----} - i \del_{--}R_{--}\right)=0~.}}
Hence, the R-current is vectorial $\partial_{++}j_{--}+\partial_{--}j_{++}=0$. In components we have
\eqn\Rmulcomp{\eqalign{
& R_{++}=j_{++}-i\theta^+ S_{+++} -i\bar\theta^+\bar S_{+++}-\theta^+\bar\theta^+ T_{++++}~,\cr
& R_{--} = j_{--} -i \theta^+ S_{+--} -i\bar \theta^+\bar S_{+--} -\theta^+\bar\theta^+ T_{++--}~,\cr
&\CT_{----}=T_{----}-\theta^+\del_{--}S_{+--}+\bar\theta^+\del_{--}\bar S_{+--}-\theta^+
\bar\theta^+\del_{--}\del_{++} j_{--}~.   }}

We couple this theory to linearized supergravity in a standard fashion. We introduce three real superfields, ${\cal H}_{++},{\cal H},{\cal H}_{----}$, and the linearized coupling takes the form
\eqn\linearizedzerotwo{\CL=\int d\theta^+ d\bar\theta^+\left(\CT_{----}{\cal H}_{++}+R_{--}{\cal H}+R_{++}{\cal H}_{----}\right)~.}
The complete super-diffeomorphism group is generated by the following transformations
\eqn\gentran{\eqalign{&{\cal H}_{++}\rightarrow {\cal H}_{++}+\left(\Lambda_{++}+\bar\Lambda_{++}\right)~, \cr
&{\cal H}\rightarrow {\cal H}+i\del_{--}\left(\Lambda_{++}-\bar\Lambda_{++}\right)+\del_{++} U_{--} ~,\cr
&{\cal H}_{----}\rightarrow {\cal H}_{----}+\del_{--} U_{--}~. }}
Above $\Lambda$ is chiral and $\bar\Lambda$ is anti-chiral. $U$ is a general multiplet.

The curvature is in the invariant chiral superfield
\eqn\invRchi{\eqalign{
\CR_{-}&=\bar D_+\left(i\del_{--}^2 {\cal H}_{++}+\del_{--}{\cal H}-\del_{++}{\cal H}_{----} \right)\cr
&=-i\left(\partial_{--}\bar\Psi_{+}-\partial_{++}\bar\Psi_{---}\right)
-{i\over 4}\theta^+ (R\sqrt{\gamma}-2iF)~.}}
Here $\Psi_{+} $ and $\Psi_{---}$ are components of the ``gravitino'' and originate from $\CH$ and $\CH_{----}$, respectively. $R$ and $F$ are the Ricci scalar and the field strength of the $U(1)$ R-gauge field, respectively.

\appendix{D}{$(2,2)$ and $(0,2)$ Supersymmetric Backgrounds in Superconformal Gauge}

In this appendix we repeat the classification of supersymmetric backgrounds of \ClossetPDA.  These authors used linearized supergravity to find the equations for supersymmetric backgrounds and then covariantized them.  Instead, we will use the superconformal gauge. This way we will not have to rely on linearized or the full nonlinear supergravity.   The point is that every $(2,2)$ background in two dimensions can be brought locally to a superconformal gauge and then all the information is contained in a chiral superfield $\Sigma$.  (More precisely, we will be using $U(1)_A$ supergravity where the conformal factor is in a chiral multiplet.  It is trivial to repeat the analysis in $U(1)_V$.)

As we said, the advantage of using this presentation is that there is no need to use supergravity.  We simply use flat space ordinary superconformal symmetry.

We will set the fermionic components of $\Sigma $ to zero; i.e.\ $\Sigma=\sigma + ia+\theta^2w$, $\bar \Sigma=\sigma - ia+\bar \theta^2 \bar w$.  We will assume that $\sigma$, which is the conformal factor, is real.  But we will allow non-unitary backgrounds in which $a$ can be complex and $\bar w$ is not the complex conjugate of $w$.

Most of our analysis will be local.  The global considerations are easily implemented later. We will denote the Killing spinors for the supersymmetry variation as $\zeta^\alpha,\ \bar\zeta^\alpha$ with $\alpha=\pm$, and will view them as four independent complex variables (no particular reality).
The conditions for supersymmetry are
\eqn\ekspinor{\eqalign{
&\partial_{\pm\pm}\zeta^\mp = \partial_{\pm\pm}\bar \zeta^{\mp }=0\cr
&\partial_{++}\left( e^{\sigma + ia}\overline{\zeta}^+\right)+{i\over 2}w e^{\sigma + ia}\zeta^-=0,\quad
\partial_{--}\left( e^{\sigma + ia}\overline{\zeta}^-\right)-{i\over 2}w e^{\sigma + ia}\zeta^+=0\cr
&\partial_{++}\left( e^{\sigma - ia}\zeta^+\right)+{i\over 2}\bar w e^{\sigma -ia}\bar\zeta^-=0,\quad
\partial_{--}\left( e^{\sigma - ia}\zeta^-\right)-{i\over 2}\bar w e^{\sigma - ia}\bar\zeta^+=0~.
}}
The first equation is the standard restriction due to flat space superconformal symmetry.  The remaining equations state that the fermionic components of $e^\Sigma$ and $e^{\bar \Sigma}$ are invariant.
Note that $\zeta^+$ and $\bar \zeta^-$ have the same $U(1)_A$ R-charge, and $e^{\sigma+ia}$ and $w$ are neutral under it.

The ${\cal N}=(2,2)$ supersymmetric background $\Sigma$ can be classified as follows according to the preserved supercharges given by the Killing spinors $\zeta^\alpha,\bar\zeta^\alpha$.

The first class of backgrounds preserves one supercharge of a given $U(1)_A$ charge. Without loss of generality we can take the Killing spinors as nonzero $(\zeta^+,\bar\zeta^-)$ and $\bar\zeta^+=\zeta^-=0$. We find for every $\sigma$ and $a$
\eqn\eqnonesusy{\eqalign{
&\Sigma=\sigma+ia -2i\theta^2 {\bar\zeta^-\over \zeta^+} \partial_{--}\left(\sigma+ia +\log\bar\zeta^-\right)\cr
&\bar\Sigma=\sigma-ia +2i\bar\theta^2 {\zeta^+\over\bar\zeta^-} \partial_{++}\left(\sigma-ia +\log \zeta^+\right)}}
where $\zeta^+=\zeta^+(x^{++})$ and $\bar\zeta^-=\bar\zeta^-(x^{--})$ are arbitrary functions of $x^{++}$ and $x^{--}$ respectively.  (In Euclidean space they are holomorphic and anti-holomorphic functions.)  Recall that our analysis is local; global considerations restrict these functions.

The second class of backgrounds preserves two supercharges with opposite $U(1)_A$ R-charge,
denoted as $(\zeta_1^+,\bar\zeta_1^-)$ and $(\bar\zeta_2^+,\zeta_2^-)$.  Imposing invariance under $(\zeta_1^+,\bar\zeta_1^-)$ we find again \eqnonesusy\ with $\zeta\to \zeta_1$.  Imposing invariance under $(\bar\zeta_2^+,\zeta_2^-)$ we find another expression, which is related to \eqnonesusy\ by $\zeta \to \bar\zeta_2$, $w\leftrightarrow\bar w$, $a \to -a$.
Consistency of the two solutions constrains $\sigma$ and $a$
\eqn\sigmacon{\left(\zeta_1^+\bar \zeta_2^+\partial_{++} + \zeta_2^-\bar \zeta_1^-\partial_{--}\right)\left(\sigma + ia + \log \bar \zeta_2^+(x^{++})\bar\zeta_1^-(x^{--})\right)=0 ~.} This means that $\sigma $ and $a$ are invariant under the vector $v\equiv \zeta_1^+\bar\zeta_2^+\partial_{++}+\zeta_2^-\bar\zeta_1^-\partial_{--}$ (up to a superconformal transformation by $\log \bar \zeta_2^+(x^{++})\bar\zeta_1^-(x^{--})$).

There are two such cases depending on whether $v=0$ or $v\neq 0$.  They lead to the topological twist and $\Omega$-deformation respectively.

For $v=0$ it suffices to assume that one component of $v$ vanishes.
Without loss of generality take $\zeta_1^+=0$ with nonzero $\bar\zeta_1^-$.
The solution \eqnonesusy\ gives
\eqn\eqntopsusy{\eqalign{
&\Sigma=\theta^2w,\quad
\bar\Sigma=2\sigma,}}
i.e.\ $a=i\sigma$, $\bar w=0$ with arbitrary $\sigma$ and $w$.
This is the topological twist. The anti-topological twist corresponds to $a=-i\sigma$.

In the second case $v\neq 0$ and $\sigma$ and $a$ must have an isometry given by the vector $v$. Up to a conformal transformation we can take the Killing spinors to be $(\zeta_1^+,\bar\zeta_1^-)=(\epsilon x^{++},1)$ and $(\bar\zeta_2^+,\zeta_2^-)=(1,-\epsilon x^{--})$. Here $\epsilon$ is a constant describing the $\Omega$-deformation.  The corresponding isometry vector is $v=\epsilon\left(x^{++}\partial_{++}-x^{--}\partial_{--}\right)$. Therefore, $\sigma$ and $a$ are arbitrary function of the invariant combination $x^{++}x^{--}$.
The supersymmetry algebra satisfies $\delta_1 \delta_2 + \delta_2 \delta_1=i{\cal L}_v\neq 0$. And the background is
\eqn\eqnwtwosusy{\eqalign{
&\Sigma=\left(\sigma+ia\right)- {2i\over \epsilon} \theta^2 {1\over x^{++} }\partial_{--}\left(\sigma+ia\right),\cr
&\bar\Sigma=\left(\sigma-ia\right)+2i\epsilon\bar\theta^2  x^{++} \partial_{++}\left(\sigma-ia+\log x^{++}\right).}}

Two limiting cases are interesting.  For $\epsilon\to 0$, the solution \eqnwtwosusy\ becomes the topological background \eqntopsusy.
For $a=i\sigma$ with nonzero $\epsilon$, equation \eqnwtwosusy\ reproduces the two-dimensional $\Omega$-background in \ClossetPDA\
\eqn\eqnomega{\eqalign{
&\Sigma=0,\quad
\bar\Sigma=2\sigma+2i\epsilon\bar\theta^2 x^{--} \partial_{--}\left(2\sigma+\log x^{--}\right).}}

The background \eqnwtwosusy\ preserves maximally four supercharges, if and only if the spacetime metric is maximally symmetric
and the $U(1)_A$ gauge field has zero curvature $\partial_{++}\partial_{--}a=0$.
For a sphere the maximally supersymmetric background is given by
\eqn\eqnsphere{\eqalign{
&\Sigma=-\log\left(1+x^2\right)+\theta^2 { 2i/ \epsilon \over 1+x^2}\cr
&\bar\Sigma=-\log\left(1+x^2\right)+\bar\theta^2   {2i\epsilon \over 1+x^2}
~,}}
where $x^2\equiv x^{++}x^{--}$.
The sphere background~\eqnsphere\  is not unitary since $\Sigma^*\neq \bar\Sigma$.
The result agrees with the supersymmetric sphere background of \refs{\BeniniUI,\DoroudXW}.

Similarly we can classify the ${\cal N}=(0,2)$ supersymmetric backgrounds.
They are determined by imposing the vanishing right-moving supersymmetry variation on the chiral superfield $e^\Sigma=e^{\sigma+ia}$
\eqn\eqnks{\eqalign{
& \partial_{--}\zeta^+ =\partial_{--}\bar\zeta^+=0~, \cr
&\partial_{++}\left(e^{\sigma+ia}\bar\zeta^+\right)=0~, \cr
&\partial_{++}\left(e^{\sigma-ia}\zeta^+\right)=0~.}}
There is only one class of ${\cal N}=(0,2)$ smooth supersymmetry backgrounds.
They are referred to as (anti-) topological half-twist in~\WittenYC\ .

\listrefs
\bye